# Transient Structures and Stream Interaction Regions in the Solar Wind: Results from EISCAT Interplanetary Scintillation, STEREO HI and *Venus Express* ASPERA-4 Measurements.

G. D. Dorrian[1], A. R. Breen[2], J. A. Davies[3], A. P. Rouillard[4], R. A. Fallows[2], I. C. Whittaker[2], D. S. Brown[5], R. A. Harrison[3], C. J. Davis[3], and M. Grande[2]

[1] *Astrophysics Research Centre, Physics Building, Queen's University of Belfast, Belfast, County Antrim, Northern Ireland, UK, BT7 1NN*

E-mail: g.dorrian@qub.ac.uk

[2] *Institute of Mathematics and Physics, Aberystwyth University, Penglais Hill. Aberystwyth, Ceredigion, Wales, UK, SY23 3BZ*

[3] *Space Science and Technology Department, Science and Technology Facilities Council, Rutherford Appleton Laboratory, Harwell Campus, Didcot, Oxfordshire, England, UK, OX11 0QX*

[4] *College of Science, George Mason University, Fairfax, USA, VA 22030*

[5] *Jeremiah Horrocks Institute, University of Central Lancashire, Preston, Lancashire, England, UK, PR1 2HE*

**Abstract** We discuss the detection and evolution of a complex series of transient and quasi-static solar wind structures in the days following the well-known comet 2P / Encke tail disconnection event in April 2007. The evolution of transient solar wind structures ranging in size from $< 10^5$ km to $> 10^6$ km was characterized using one-minute time resolution observation of Interplanetary Scintillation (IPS) made using the European Incoherent SCATter (EISCAT) radar system. Simultaneously, the global structure and evolution of these features was characterized by the Heliospheric Imagers (HI) on the Solar TERrestrial RElations Observatory (STEREO) spacecraft, placing the IPS observations in context. Of particular interest was the observation of one transient in the slow wind apparently being swept up and entrained by a Stream Interaction Region (SIR). The SIR itself was later detected *in-situ* at Venus by the Analyser of Space Plasma and Energetic Atoms (ASPERA-4) instrument on the *Venus Express* (VEX) spacecraft. The availability of such diverse data sources over a range of different time resolutions enables us to develop a global picture of these complex events that would not have been possible if these instruments were used in isolation. We suggest that the range of solar wind transients discussed here maybe the interplanetary counterparts of transient structures previously reported from coronagraph observations and are likely to correspond to transient magnetic structures reported in *in-situ* measurements in interplanetary space. The results reported here also provide the first indication of heliocentric distances at which transients become entrained.

**Keywords:** *Magnetic fields, Interplanetary; Radio Scintillation; Solar Wind, Disturbances, Stream Interaction Region*

## 1. Introduction

Interplanetary Scintillation (IPS) is the rapid amplitude variation induced by solar wind irregularities on received radio emission. Observations of IPS on astronomical radio source signals has been used as a solar wind flow tracer for more than 45 years (e.g. Hewish, Scott, and Wills, 1964; Bourgois 1972; Watanabe, *et al.*, 1973; Bourgois, *et al.*, 1985; Coles, 1995, Breen, *et al.*, 1996a, 1996b; Asai, *et al.*, 1998; Fallows, *et al.*, 2006; Dorrian, *et al.*, 2008).

If the radio source is observed with two antennas, and the orientation of the projection into the plane-of-sky of their baseline is close to parallel to the outflow axis of the solar wind, then the amplitude variation patterns received at the two antennas will be significantly correlated (Armstrong and Coles, 1972). The maximum cross-correlation occurs at a time lag that depends on the velocity of the solar wind irregularities and the radial component of the projection of the antenna separation into the plane-of-sky (e.g. Armstrong and Coles, 1972; Bourgois, *et al.*, 1985). The geometry of such an observation is shown schematically in Figure 1. The accuracy with which the velocities of multiple streams crossing the ray path can be determined improves as the projected baseline length increases (e.g. Coles, 1996; Klinglesmith, 1997; Breen, *et al.*, 2006).

More detailed analysis of two-site long-baseline observations of IPS can provide further information, such as the spread of velocities in a single solar wind stream and the random variation in transverse velocity, and can also provide improved estimates of bulk outflow velocity (e.g. Coles, 1996; Fallows, Breen, and Dorrian, 2008). Such analysis relies on supplementary information, for example from white-light coronal maps derived from coronagraph data. This information is used to identify the location in the ray path of regions of fast and slow solar wind so that they can be fitted using a weak-scattering model (Coles, 1996). The compression region of stream interaction regions (SIRs) can be identified in observations of IPS as regions of enhanced scintillation level. Additionally, they often possess a velocity intermediate between that of fast and slow streams, and overlie the leading boundary of coronal holes, as seen in coronal maps (Klinglesmith, 1997; Breen, *et al.*, 1998, 1999; Bisi, *et al.*, 2010).

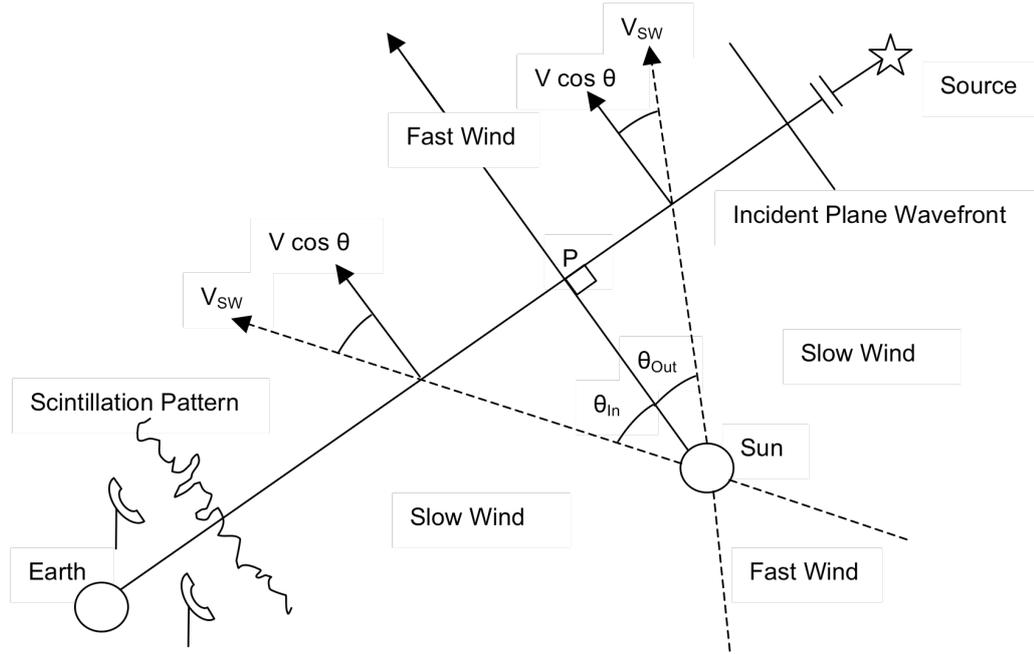

**Figure 1.** The geometry of two-station IPS observations. $\theta_{in}$ and $\theta_{out}$ are the angles formed between a line from the Sun through the point of closest approach of the ray path (the P-point) and the Earthward and source ward fast/slow wind boundaries, respectively. These angles are used as parameters in the IPS weak scattering model, discussed in Section 2.2. The arrows labeled V cosθ indicate that the technique is sensitive to the plane-of-sky component of solar wind velocity ($V_{PoS}$).

The presence of a Coronal Mass Ejection (CME) in the ray path can be recognised by the appearance of some or all of the following signatures (Canals, *et al.*, 2002; Canals, 2002; Jones, *et al.*, 2007; Dorrian, *et al.*, 2008).

1. A rapid variation (on the order of hours or less) of the solar wind velocity;
2. A negative lobe in the cross-correlation function near zero time-lag, produced by a rotation of the interplanetary magnetic field at some point in the ray path (Klinglesmith, Grall, and Coles, 1996);
3. A rapid variation (on the order of hours or less) in the shape of the cross-correlation function;
4. A rapid increase and variation in scintillation level on timescales of hours or less.

As dense CME-related structures crossing the IPS ray path increase the scintillation level, this reduces the integration time necessary to acquire well-defined spectra. As a result, either a higher time resolution analysis can be performed on the IPS data, or more degrees of fitting are available when attempting to model scattering along the ray path (See Section 2.2).

The ultra-high frequency (UHF) mainland European Incoherent SCATter (EISCAT: Rishbeth and Williams, 1985) mainland radar consists of three fully steerable 32-m parabolic antennas, sited near Tromsø (Norway), Kiruna (Sweden), and Sodankylä (Finland). When operating in IPS mode, the Kiruna and Sodankylä antennas receive on a 5.4 MHz wide band centred at 1420 MHz (see Wannberg, *et al.*, 2002 for details on the 1420 MHz system upgrade), while the Tromsø site receives on a 5.4 MHz band centred at 928 MHz. Received signals are sampled at $10^4$ Hz and are subsequently binned to yield measurements of source intensity at 0.01s resolution (Fallows, Breen, and Dorrian, 2008). Analysis has shown that dual-frequency observations centred on 1420 and 928 MHz can be combined to provide 2-site observations of solar wind velocity, over a wide range of heliocentric distances. For example Fallows, *et al*. (2006) have made succesful solar wind velocity measurements by observing IPS using radio sources with plane-of-sky heliocentric distances ranging from 27 $R_\odot$ to 85 $R_\odot$

EISCAT was first used for observing IPS in the early 1980s (Bourgois, *et al.,* 1985). It offers the advantage of relatively long baselines for two-site observations (up to 390 km parallel baseline, e.g. Breen, 1996a) which allows fast and slow streams of solar wind to be resolved (e.g. Grall *et al.*, 1996; Breen *et al.*, 1996b).

Observations of Thomson scattering of photospheric light off solar wind structures was pioneered in the 1970s with the use of photometers carried aboard the *Helios-1* and *-2* spacecraft (e.g. Jackson, 1985). The technique has been developed in recent years with the advent of instruments such as the Solar Mass Ejection Imager (SMEI: Eyles *et al*., 2003; Jackson *et al*., 2004) aboard the *Coiolis* spacecraft, in Earth orbit, and the Heliospheric Imagers (HI: Eyles *et al*., 2009; Harrison *et al*., 2009.) aboard the Solar TERrestrial RElations Observatory (STEREO: Kaiser, 2005; Kaiser *et al*., 2008).

STEREO consists of a pair of spacecraft in heliocentric orbit at approximately 1 AU. STEREO-A, being at a slightly closer distance to the Sun than 1 AU, orbits ahead of Earth. STEREO-B, at a slightly greater distance than 1 AU, lags behind Earth in its orbit. The separation angle between each spacecraft and the Earth increases by 22° per year. In this paper we use data only from STEREO-A.

The HI instruments on each STEREO spacecraft form part of the Solar Earth Connection Coronal and Heliospheric Imager (SECCHI) package (Howard *et al*., 2008). Each HI instrument comprises of two wide field, white-light cameras (HI-1 and HI-2), designed principally to observe the propagation of Earth bound CMEs (Eyles *et al*. 2009). The cameras are shielded from stray light by a series of baffles and are capable of imaging features in the field-of-view down to $3\times10^{-15}$ and $3\times10^{-16}$ solar brightnesses for HI-1 and HI-2, respectively. HI-1 has a field-of-view of 20°, centered at 14° elongation from solar disk center and an image cadence of forty minutes. HI-2 has a field-of-view of 70° centered at 54° from disk center and an image cadence of one hundred and twenty minutes. Note that the fields-of-view of the two cameras overlap. HI coverage of the inner heliosphere thus extends from within 4° of solar disk center out to 1 AU and beyond. Such coverage enables uninterrupted monitoring of the propagation of any Earth-bound solar wind phenomena all the way from the Sun to Earth.

Any variation in electron density due to the presence of transient features such as CMEs will be detected by HI as a consequent variation in received white-light intensity. Rouillard *et al.* (2009) reported one of the first detections by HI of a solar transient entrained by a SIR. The event reported by those authors is very similar in nature to the one we discuss here. In both cases a SIR entrains a transient sometime after leaving the lower corona. Although Rouillard *et al.* (2009) established that the transient they studied had become entrained by the time it entered the field-of-view of HI-2, they could not place any constraints on where and when the transient had become entrained. In this study we combine observations of IPS and white-light in the inner heliosphere, which enable us to place some minimum heliocentric distance on the entrainment of the transient by the SIR.

In this study, the availability of IPS observations covering the inner heliosphere (i.e. within the field-of-view of HI-1 on STEREO-A) have allowed us to characterize both the velocity of the transient, and that of the SIR, simultaneously. A velocity of the transient slower than that of the SIR would indicate that the SIR had not yet swept up the transient.

Observations both of IPS and Thomson-scattered white light are path-integrated, with maximum scattering biased towards the region of the line-of-sight closest to the Sun, known as the point of closest approach (P-point, or "impact parameter"). The weighting functions are not identical, however, and this leads to significant differences in the way that a three-dimensional (3-D) solar wind structure is detected by these two techniques. The Thomson scattering effects, which affect white-light observations remain small, however, out to approximately 100 $R_\odot$ (Vourlidas and Howard, 2006).

*In-situ* measurements of the solar wind inside Earth orbit were first made by the Mariner-2 spacecraft in 1962 (Neugebauer and Snyder, 1966). The *Helios* missions made *in-situ* measurements as close to the Sun as 65 solar radii ($R_\odot$), equivalent to 0.3 AU (Marsch *et al.*, 1982). Currently *in-situ* measurements inside 1 AU are available from the *Venus Express* spacecraft (VEX: Svedhem *et al.*, 2007) and the MErcury Surface, Space ENvironment, GEochemistry and Ranging spacecraft (MESSENGER: Slavin *et al.*, 2009).

VEX was launched in November 2005 and arrived at Venus in April 2006. It is currently in a highly elliptical orbit, with a period of 24 hours and a periapsis approximately 250 km above the planet's surface. The ASPERA-4 (Analyser of Space Plasma and EneRgetic Atoms: Barabash *et al.*, 2007) instrument package is comprised of four instruments; the Neutral Particle Detector, Neutral Particle Imager, Electron Spectrometer, and the Ion Mass Analyser (IMA) that is mounted separately. Together this package makes *in-situ* measurements of both the ionized and neutral particle environment.

The IMA instrument, data from which is used in this study, is a top hat electrostatic analyser, which detects ions in the range of 10 eV to 30 keV, with mass distinction. The azimuthal angular coverage is 360°, divided into 16 sectors sampled simultaneously. The instrument sequentially samples eight polar bins, these bins covering a total range in polar angle of -45° to 45°. The accumulation time for a full spectrum comprising all angles and energies is 192 seconds. The instrument takes observations twice during each orbit, for 120 minutes at periapsis, and for 60 minutes at apoapsis. The periaptic activation records data within Venus's magnetosphere and bow shock, whereas the apoaptic section records purely solar wind data. Ion count and average energy are indicative of ion density and velocity, respectively. SIRs are identified in by a rapid increase in ion count and energy, indicative of increased ion density and velocity, respectively, followed by a reduction in average energy over several consecutive days (e.g. Gosling *et al.*, 1978). This profile corresponds to the instrument encountering the compressed material at the leading edge of the SIR, followed by the extended rarefaction region, which is dominated by higher energy, fast solar wind particles.

In this paper, we discuss observations of solar wind transients of different scale sizes, which we categorize by how long their signatures exist for in the data collected from an IPS observing run.

Solar wind transients that generate the IPS at the frequencies used here have scale sizes of the order of 100 km. We refer to these here as "micro-scale transients" and they occupy the ray path for time scales of the order of seconds.

We use the term "small-scale transients" for those which occupy the IPS ray path for less than 30 minutes whilst travelling at a typical slow wind velocity of 350 km s$^{-1}$, implying scale sizes of less than approximately 600000 km. The transients observed on 21-24 April 2007, using IPS, which will be discussed later, have typical scales in the 50000-70000 km range. Small-scale transients may be the same phenomena as the pixel brightening features reported by Tappin, Simnett, and Lyons (1999), which were used by those authors to assess the velocity of the slow solar wind.

We use the term "meso-scale transients" for features such as those observed on 25 April 2007, which occupy the ray path for between 30 and 60 minutes but are nonetheless considerably smaller than CMEs. Travelling at 350 km s$^{-1}$, a meso-scale transient would therefore have a scale size ranging from approximately 600000-1200000 km.

Rouillard *et al.* (2009) reported on the relationship between meso-scale transient and a SIR, observed in July 2007, interpreting the transient as having originated as a slow wind structure. In this study we build on their work by adding an estimate of the minimum heliocentric distance at which such entrainment occurs.

Borovsky (2006) discusses turbulent eddies detected in solar wind flow of comparable scale sizes to the meso-scale transients we discuss here. The extensive *in-situ* OMNI solar wind data set he uses is collected at 1 AU. The observations of IPS reported in this paper were made whilst the ray path lay at a minimum heliocentric distance of ~0.3 AU, and hence confirm the presence of such structure at heliocentric distances significantly less than 1 AU. However, we cannot confirm here whether the transient structures underwent reconnection prior to detection, as suggested by Borovsky (2008).

Meso-scale transients may also be the same phenomena as the well-known "Sheeley blobs" (Sheeley *et al.*, 1997). Smaller structures may correspond to pixel brightening as reported by Tappin, Simnett, and Lyons (1999). The largest scale transients are CMEs, which occupy a significant proportion of the heliosphere with a total angular extent of several tens of degrees and are present in the ray path for at least several hours.

## 2. Observations

### 2.1 Observations of Interplanetary Scintillation using EISCAT

A series of observations of IPS using radio source J0318+164, made with the EISCAT Kiruna and Sodankylä antennas, started on 21 April 2007. At this time the point of closest approach of the ray path from the radio source to the antennas lay at a heliographic latitude of 11.0º S and a heliocentric distance of 80.9 R$_\odot$. Between 30 and 60 minutes of data was recorded each day on 21, 22, 24 and 25 April, with observations from 25 April being of particular interest. A summary of the IPS observations is given in Table 1.

The IPS data were analysed in two ways. In the first method, a ten-minute sliding window, advanced in one-minute increments, was used to analyze the cross-correlation function. Thus, the first time bin for analysis on 21 April started at 1515 UT and included all data up to 1525 UT, the second time bin began at 1516 UT and extended to 1526 UT, and so on up to the last available ten minutes. Estimates of the projection of solar wind velocity into the plane-of-sky (V$_{PoS}$) were subsequently obtained directly from the parallel baseline length of the IPS observation and the time lag of the peak cross-correlation. This method provides good temporal resolution for detecting variations in bulk flow speed.

| Date | Time (UT) | Observation Length (min) | Antenna Pair | Heliographic Latitude | Carrington Longitude | P-point to Sun distance ($R_\odot$) |
|---|---|---|---|---|---|---|
| 21 April | 15:15 | 30 | KS | -11.3 | 33.0 | 76.0 |
| 22 April | 15:15 | 30 | KS | -11.6 | 17.2 | 72.6 |
| 24 April | 14:45 | 60 | KS | -12.1 | 347.5 | 65.8 |
| 25 April | 14:30 | 60 | KS | -12.5 | 332.3 | 62.4 |

**Table 1**. IPS observation parameters for 21-25 April 2007. K denotes Kiruna and S, Sodankylä. The heliographic latitude and Carrington longitude positions given are for the P-point.

In the second method, which will be discussed in Section 2.2, the data were fitted with the weak scattering model. Ten-minute integration times were not long enough to allow sufficiently well defined IPS spectra to develop such that multiple solar wind streams could be accurately fitted with the weak scattering model. thirty-minute integration times were, therefore, used for this stage of the analysis. This meant sacrificing some temporal resolution in favour of the ability to model multiple solar wind streams passing across the ray path.

In the first analysis method, each ten-minute time bin was analysed to provide an estimate of velocity from the time lag at which maximum cross-correlation occurred ($V_{PoS}$). The cross-correlation functions and velocities displayed in Figures 2-5 below are therefore ten-minute averages. For each of the ten-minute bins, the lower and upper thresholds, respectively, of the high- and low-pass filters applied to the power spectrum (e.g. Fallows, Breen, and Dorrian 2008) were maintained at constant values of 0.15 Hz and 4 Hz, respectively. This was to ensure that any variation in the velocities and/or scintillation levels returned by the analysis programme were not due to artificial changes in the analysis routine.
The presence of highly disturbed solar wind conditions from 21 to 25 April 2007 was revealed by observations of IPS. Rapid variations in $V_{PoS}$ were detected that were accompanied, in particular during the first three days of observations, by frequent short-lived Interplanetary Magnetic Field (IMF) rotations in the ray path. Significant levels of anti-correlation, appearing as negative lobes in the cross-correlation function, reveal such field rotations. The short-lived nature of these changes was deduced from the fact that, over the course of a single set of observations of ten- or sixty-minute duration, the anti-correlation in the cross-correlation function would frequently come and go. The $V_{PoS}$ throughout each period of observation, as well as sample cross-correlation functions, from data recorded on 21, 22, 24 and 25 April are shown in Figures 2-5.
Anti-correlation was observed in the cross-correlation function throughout the entire sixty minutes of IPS observations on 25 April, indicating that a larger, meso-scale transient was moving across the ray path. The fact that the peak cross-correlation (> 60%) was also significantly higher for observations of IPS on 25 April than on previous days, and that the area under the auto-spectra for each of the two receiver sites was also significantly greater, both imply an increase in electron density in the scattering region. Furthermore, the fact that the scintillation levels on that day were approximately an order of magnitude higher than on the previous days also reinforces this idea.

A meso-scale transient associated with increased white-light intensity (solar wind density) was observed by HI-1 on STEREO-A (HI-1A) on 25 April 2007. It will be later shown that this transient was in close proximity to the IPS ray path. The feature was latitudinally extended and structurally stable whilst it propagated anti-sunward across the HI-1A field-of-view. However, HI-2A observations suggest that it evolved significantly over the ensuing days. The HI observations of this feature are described in Section 2.3.

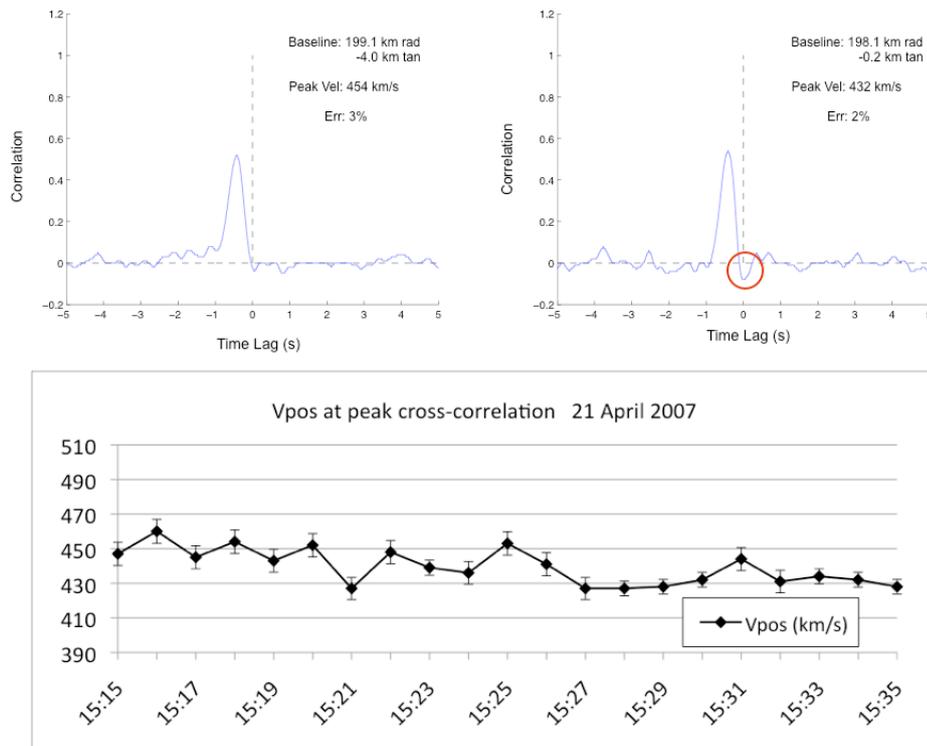

**Figure 2.** Sample cross-correlation functions (top left and right panels) from observations of IPS on 21 April 2007, one showing the appearance of anti-correlation (circled) near zero time lag, indicating the presence of magnetic field rotation in the ray path. Each cross-correlation function is produced from a ten-minute data bin. Field rotations are observed on timescales as short as 15 minutes, suggesting structures with scale sizes of less than 400000 km, placing them in the small-scale transient category. The bottom plot shows the variation in $V_{PoS}$ recorded over the course of the observing run (times indicate the start of each ten-minute bin). Solar wind velocity throughout this paper is given in km s$^{-1}$.

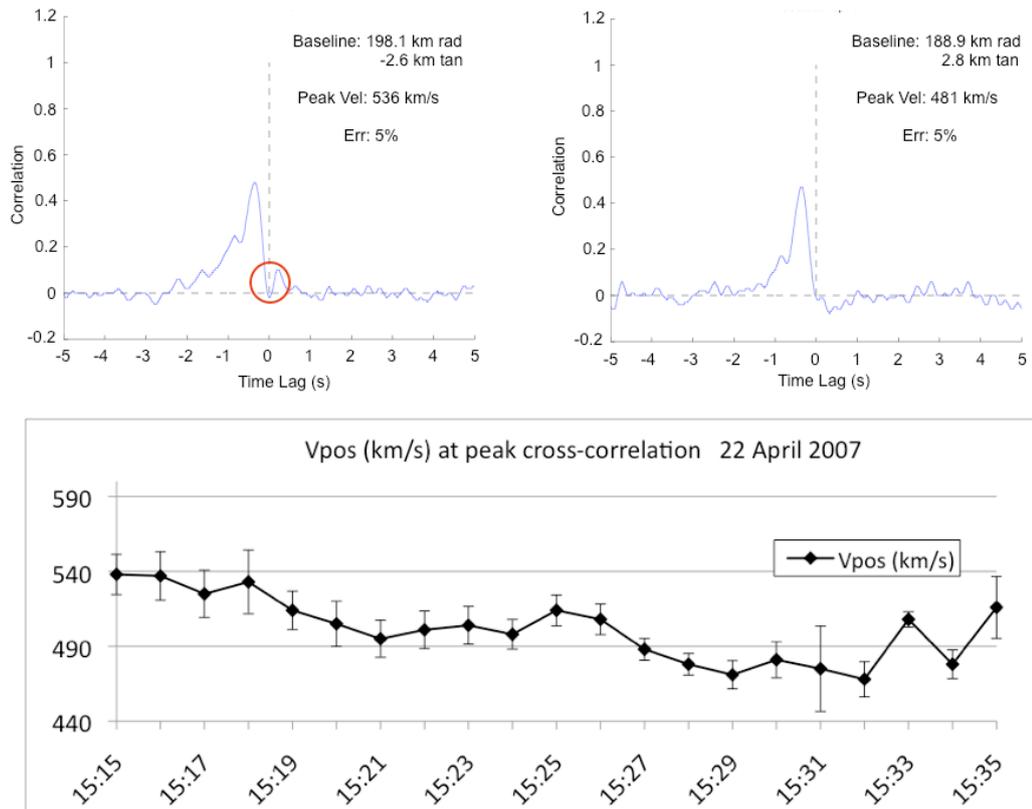

**Figure 3.** Sample cross-correlation functions (top left and right panels) from observations of IPS on 22 April 2007. Anti-correlation near zero time lag is visible in the top left (circled) panel corresponding to the ten-minute bin beginning at 15:16 UT. By the time of the bin beginning at 15:25 UT (top right), this anti-correlation has disappeared. As with observations of IPS on 21 April, this is indicative of small-scale transients crossing the ray path. The bottom plot shows the variation in $V_{PoS}$ over the course of the observing run.

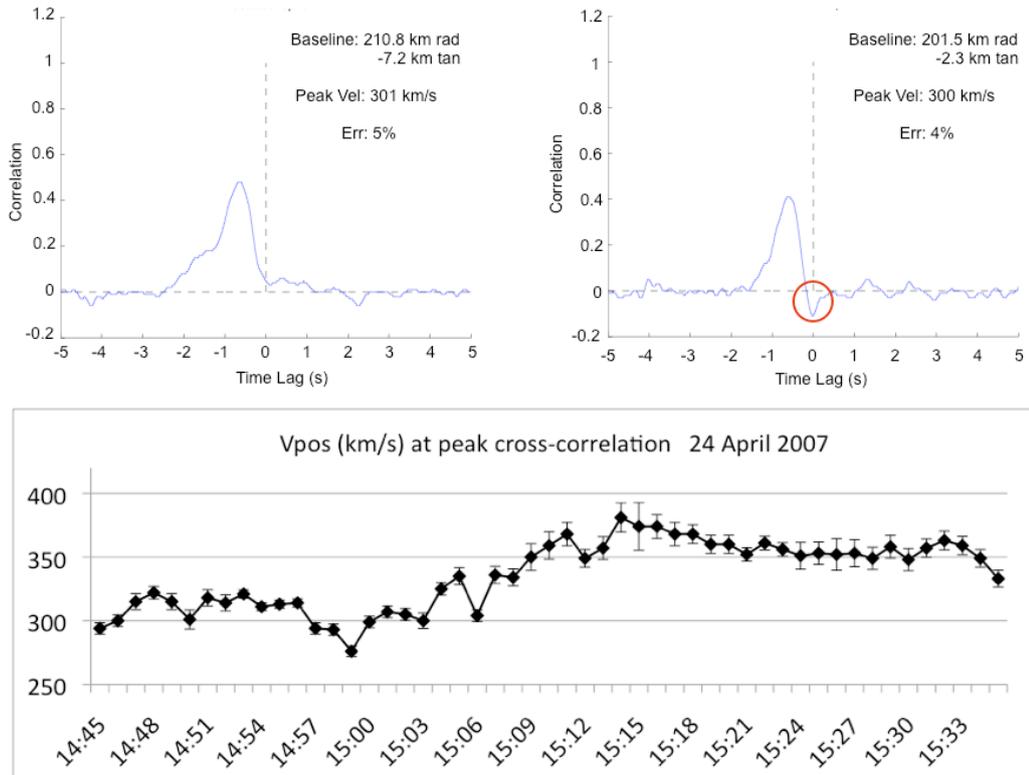

**Figure 4.** Sample cross-correlation functions (top left and right panels) from observations of IPS on 24 April 2007. There is no anti-correlation near zero time-lag visible in the top left panel corresponding to the ten-minute bin beginning at 14:50 UT. However by the time of the bin beginning at 15:02 UT, there is clear anti-correlation (circled). The bottom plot shows the variation in $V_{PoS}$ recorded over the course of the observing run; a clear increase in solar wind velocity is detected from 15:03 UT onwards, corresponding with the appearance of anti-correlation.

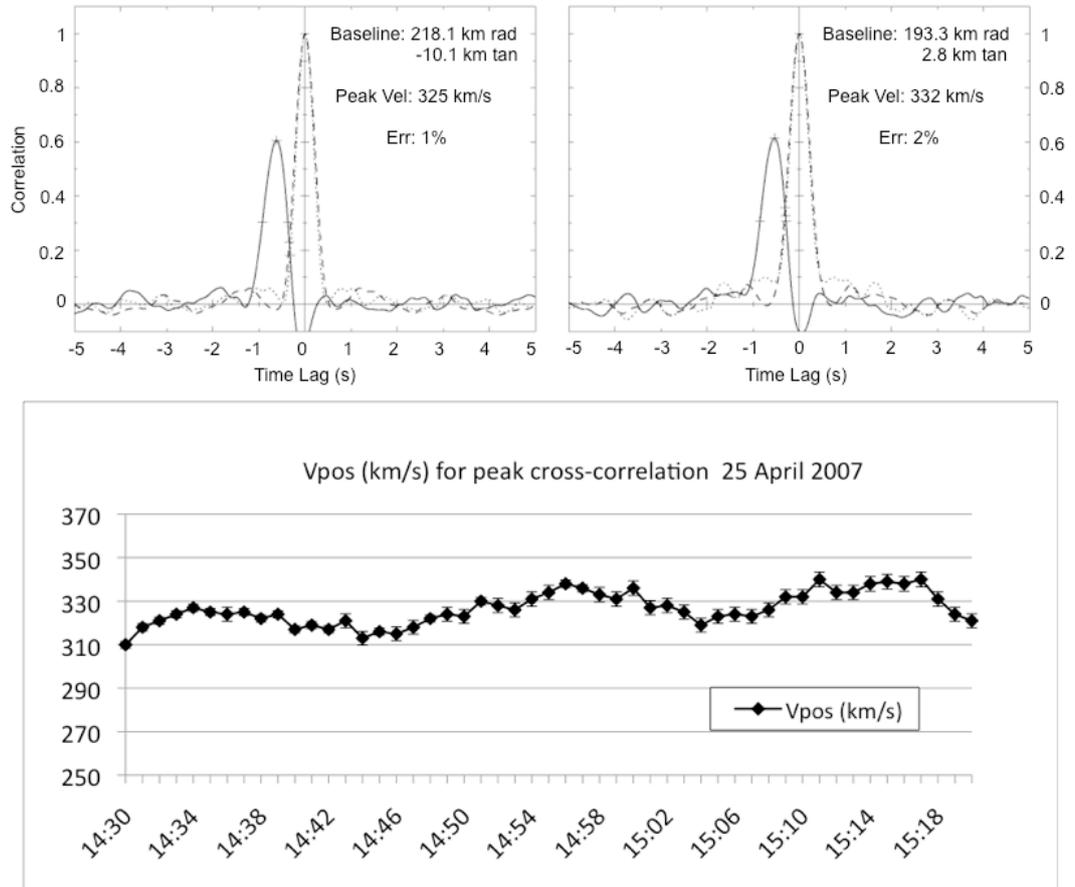

**Figure 5.** The top panels show sample cross-correlation functions (solid lines) obtained from the IPS data taken on 25 April. The plot at top left is from the time bin starting at 14:35 UT and the plot at top right is from 15:10 UT. Anti-correlation was present throughout the full sixty-minute observation, indicating that a meso-scale feature was occupying some portion of the ray path. Based on its velocity of approximately 330 km s$^{-1}$, the feature had a minimum scale size of 1180000 km. The bottom plot shows $V_{PoS}$ recorded over the course of the sixty-minute observation of IPS on 25 April. A periodic variation in the velocity is evident, superposed on a general upwards trend. Such rapid variation in $V_{PoS}$ is indicative of transient activity in the ray path.

## 2.2 Modelling the IPS observations

The auto- and cross-spectral data obtained from observations of IPS on 25 April were fitted using a three-mode, two-dimensional (2-D) weak-scattering model (Bisi *et al.*, 2007; Fallows, Breen, and Dorrian, 2008). The ray path for J0318+164 on that date was ballistically mapped back along the Parker spiral to a height of 2.5 R$_\odot$. In order to assess which parts of the ray path were occupied by specific structures, the mapped ray path was then overlaid onto the appropriate Carrington map of coronal white-light intensities derived from LASCO data using tomographic inversion (Morgan, Habbal, and Lugaz, 2009), as shown in Figure 6. The dashed line on Figure 6 marks the position of the Heliospheric Current Sheet (HCS), as estimated using the technique of potential field source surface extrapolation (Altschuler & Newkirk 1969; Schatten, Wilcox, and Ness, 1969; Wang & Sheeley 1992).

The portions of the ray path embedded in fast solar wind flow were found to lie between 65° and 85° sourceward and from 50° to 90° Earthward of the P-point. Slow solar wind was modelled as occupying the portion of the ray path in between 50° Earthward to 65° sourceward. A portion of the ray path between 35° and 60° sourceward of the P-point is found to overlay the Western latitudinal boundary region between an equatorial coronal hole and denser coronal structure (circled in Figure 6). This portion was modelled as being the location of the compression region on the leading edge of the SIR.

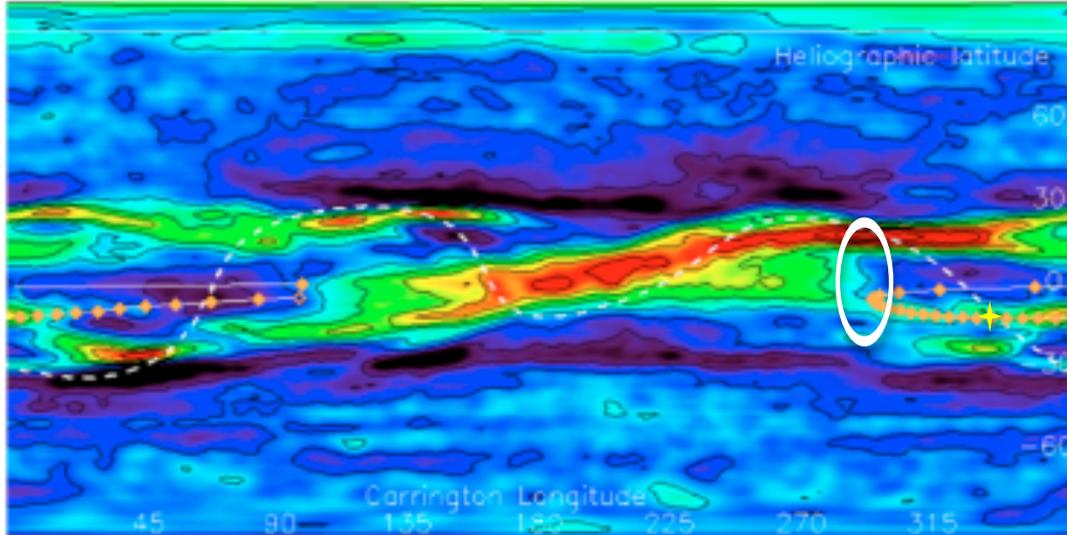

**Figure 6.** The ray path for the observation of IPS, using radio source J0318+164 on 25 April 2007, was ballistically mapped back along the Parker spiral onto a tomographic inversion map of the corona at 2.5 R$_\odot$. The tomographic map is assembled from LASCO data (Morgan *et al.*, 2009) and the white dashed line represents the position of the HCS. Coronal density is scaled with red representing the densest structures and blue the most rarefied. The solid white line indicates the mapped IPS ray path, with each orange diamond along the ray path marking 5° of extent. The diamond overlaid by a yellow star is the P-point location and the diamond with a black center shows the Earth end of the ray path. The portion of the ray path extending from 35° to 60° sourceward from the P-point (highlighted with white oval) is seen to overlie the latitudinal boundary of an equatorial coronal hole which corresponds to the location of the compression region of the SIR.

Figure 7 shows the results of fitting data obtained during the entire sixty-minute observation of IPS on 25 April. The observed cross-correlation functions (lower two solid lines) are well reproduced by model fitting (dashed lines). As discussed in Section 2.1, in order to reduce the random error in the fitting the acquired data were binned into two sequential thirty-minute data bins; results from the first and second of these two thirty-minute bins correspond to the central and lower cross-correlation functions in Figure 7, respectively. The values for the modelled velocity and its associated reduced $\chi^2$ error quoted below are derived from the full 60 minutes. The top profile in Figure 7 corresponds to the fitted auto-correlation function.

The modeled velocity was arrived at after an iterative series of trial runs, each producing an estimate of the reduced $\chi^2$ error. This was performed by using an automated function in the model program. In each trial run, the model parameter values are stepped from some user specified minimum to maximum value and with a user specified step value. For example the program can be set to run through the trials stepping from a minimum value for solar wind velocity for the compression region of 200 km s$^{-1}$ up to a maximum of 700 km s$^{-1}$ in +10 km s$^{-1}$ increments. Each run outputs the error and the corresponding model parameter values for that run. The parameter inputs used in the final model fit were those which returned the smallest, reduced $\chi^2$ error during the trials.

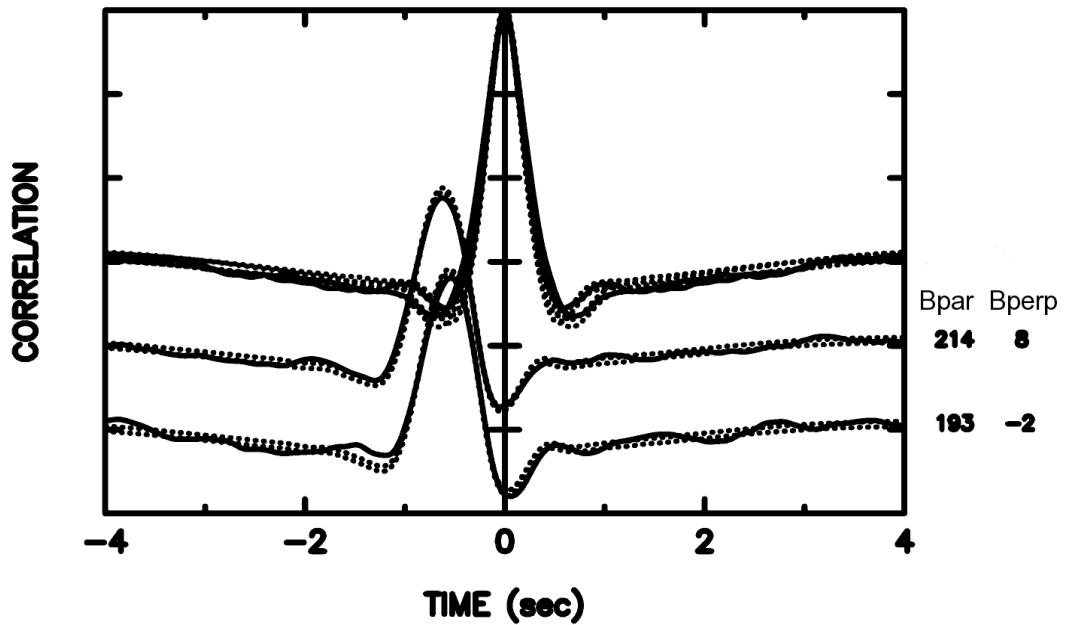

**Figure 7.** The solid lines are the auto- and cross-correlation functions obtained from the actual data and the dotted lines represent the bounds of the weak scattering model fit. The upper profile is the auto-correlation function, the middle profile corresponds to the first thirty-minute data bin (from 14:30 to 15:00 UT) and the lower profile corresponds to the second thirty-minute data bin (15:00 to 15:30 UT). It can be seen that in all cases the model well reproduces the form of the cross- and auto-correlation functions, including the near zero time lag anti-correlation, signifying the presence of the transient.

The good match between the observed and fitted correlation functions implies a successful fit with the three-stream model. The reduced $\chi^2$ error returned by the best fit was 3.93 and the correlation coefficients for the modelled cross-correlation functions were 0.62 and 0.63 for the first and second thirty-minute data bins, respectively. The model estimates for outflow velocities were 333±51 km s$^{-1}$ for the slow solar wind, 392±50 km s$^{-1}$ for the compression region of the SIR and 650±50 km s$^{-1}$ for the fast solar wind. The errors in velocity are obtained from the modeled deviation in radial flow speed across the stream (e.g. Breen *et al.*, 1999). Attempts made to separate the velocity of the transient from the velocity of the slow solar wind proved unsuccessful. The SIR velocity lies above the upper bounds of the slow solar wind implying that the transient was moving with the slow solar wind and was not yet entrained by the SIR. The position of the IPS ray path therefore places a constraint on the earliest time and minimum heliocentric distance at which entrainment occurred.

The density of the slow wind required for the successful fit, as a model parameter, is higher than would normally have been expected for a purely slow wind stream. We find that the micro-scale transients which give rise to IPS are not elongated in the direction of the interplanetary magnetic field. The best fit gives a value for the axial ratio of slow wind micro-scale irregularities of 1. This suggests the region fitted as "slow wind" contains a mix of unperturbed Parker-spiral interplanetary magnetic field and a region of significant field rotation with micro-scale transients, giving rise to IPS, elongated along the field lines (Grall *et al.*, 1997). Both of these features are also consistent with the transient lying in the slow wind (Klinglesmith, 1997).

## 2.3 Heliospheric Imager observations and geometry

An overview of the geometry is shown in Figures 8 and 9, the position of P' in the HI-1A field-of-view can be seen in Figure 10, and the evolution of the transient in the HI-2A field-of-view is shown in Figure 11. The central axis of motion of the transient in the ecliptic, as obtained from the HI event list (http://www.sstd.rl.ac.uk/Stereo/HIEventList.html), is at 89±15° heliographic longitude. Whilst this would make it possible that the transient itself would pass through P and P' (e.g. Figure 8), we do not suggest that this actually happened. Instead we simply state that a transient would cross the IPS ray path very close to the time that it passed through the HI-1A line-of-sight at P' due to their close proximity. This is clearly confirmed later by the J-maps presented in Figure 12; at the time that the transient is detected in the observations of IPS it indeed passes over the position of P' in the HI-1A field-of-view. We can therefore relate directly the transient signatures observed in IPS on 25 April, to those in HI-1A.

The apparent position of the IPS P-point in the HI-1A field-of-view (P'), at 14:50 UT on 25 April, was calculated using 3-D vector analysis. This time corresponds to the meso-scale transient crossing the IPS ray path (see Figure 10). The relative locations of STEREO-A, the transient, and Earth do not change significantly over the course of the forty-minute HI-1A exposure.

The angular separation between the STEREO-A spacecraft and Earth was only 3.78° at this time, so the view of the radio source position relative to the Sun, from STEREO-A would not be greatly different to the view from Earth. Nonetheless we performed a full correction for viewing direction.

The primary feature of interest seen in HI-1A difference images over the interval covered by Figure 10 is the transient. Over the course of the later images of the sequence this feature becomes increasingly compressed along the horizontal component of its axis of motion. The tail of comet 2P / Encke can also be seen undergoing significant dynamics at this time as discussed by Clover *et al*. (2010). A brief comparison of the velocities from that study and this one can be found in Section 4.

Over the next few days the compression region of the SIR continued to develop. The transient, now in the HI-2A field-of-view, becomes further compressed and it is at this stage that we surmise that the transient becomes entrained by the SIR (Figure 11). The radial velocity of the transient, in the ecliptic plane, according to the HI event list was 273 ±53 km s$^{-1}$; and this value lies within the error bounds of the slow solar wind velocity as obtained from the weak scattering model (333±51 km s$^{-1}$).

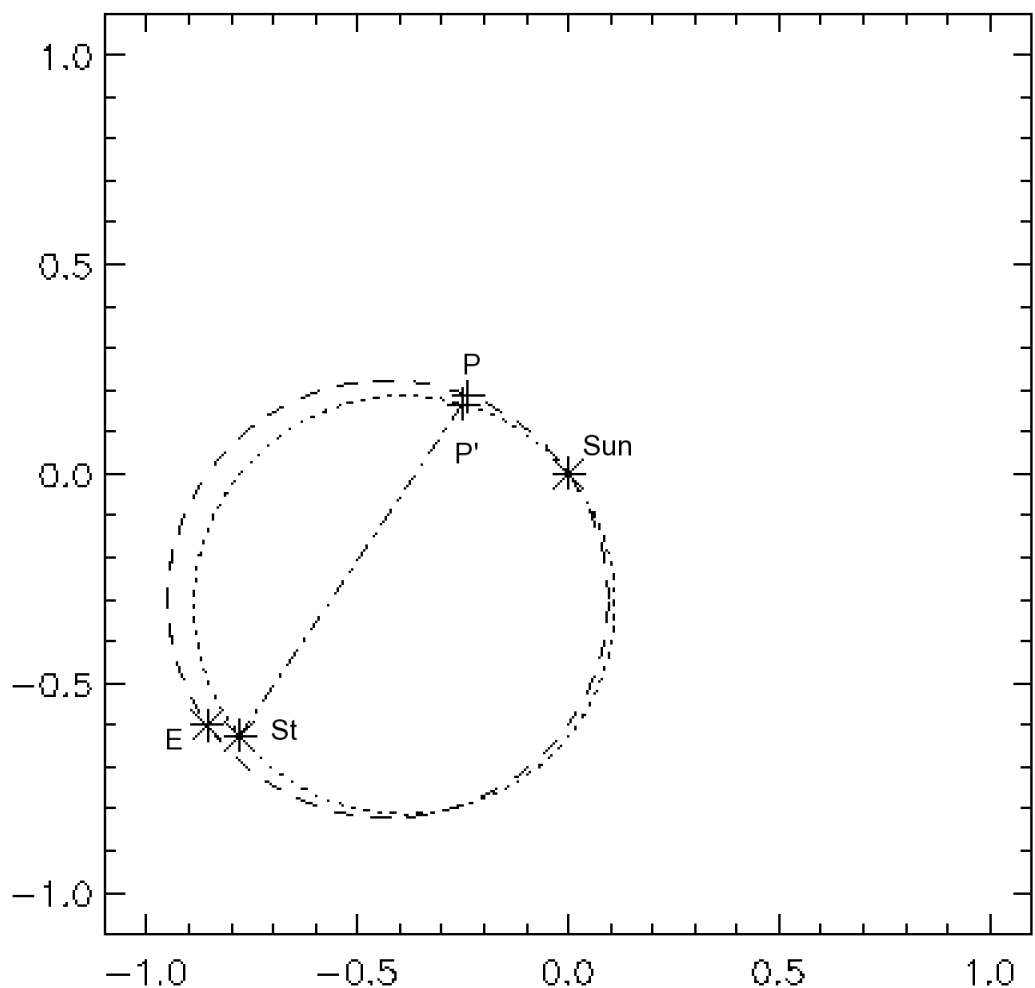

**Figure 8.** A heliocentric view showing the location of P and P', the Earth (E), STEREO-A (St) and the Sun on 25 April 2007 at 14:50 UT, corresponding with the IPS observing run. The axes are in astronomical units and the plot is plane parallel with the ecliptic. The Thomson spheres for Earth (dashed circle) and STEREO-A (dotted circle) are overlaid demonstrating that P and P' both lie on their corresponding sphere. The line-of-sight from STEREO-A to P' is also shown. The small angular separation between the Earth and the STEREO-A spacecraft at the time of the observations ensures that P and P' lie in close proximity to each other.

In 40 minutes the field-of-view of HI-1A rotates (due to spacecraft motion around the Sun) by approximately 0.03° or approximately 1.5 pixels (as the HI-1A field-of-view is 20° and 1024 pixels wide). . This is therefore of negligible consequence for the line-of-sight corrections used here.

A J-map (e.g. Davies *et al.*, 2009) can be constructed from HI data at a position angle that encompasses P', providing a useful tool for investigating what structures are seen to cross the ray path at the time and location of the IPS observation. The time-elongation profile of a solar wind transient extracted from such a J-map can also be used to estimate its radial velocity and direction of propagation (e.g. Rouillard *et al.*, 2008).

In Figure 12 two J-maps are presented; the top one is comprised from running difference HI-1A and HI-2A images and the bottom one from 24 hour background subtracted HI-1A images, over the period under discussion. A 24-hour running background, centered at the time of each image is used to remove stable solar wind structure, such as coronal streamers, and the F-corona. Hence, only the rapidly evolving transient material remains visible in the images. In this lower panel, the position of P' for the observation of J0318+164 on 24 April is also shown to contrast the solar wind conditions at the location of P' on 25 April. Feature A is the meso-scale transient that travels from the inner edge of the HI-1A field-of-view to the position of P' on 25 April. As noted before, the true radial velocity of this feature, as obtained from the STEREO HI event list is 273±53 km s$^{-1}$, comparable to the slow wind radial velocity from the IPS model fitting. It is this feature that is interpreted as causing the anti-correlation observed in the observations of IPS on 25 April.

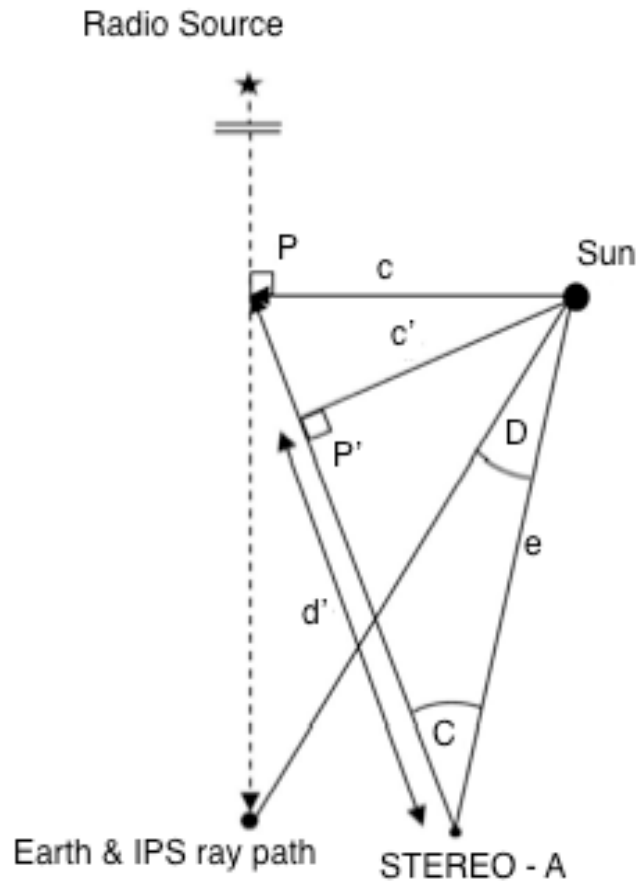

**Figure 9.** The general geometrical layout of STEREO-A, Earth, P and P', as viewed from above, during the joint IPS/HI observation on 25 April 2007. The angles C and D were calculated to be 17.52° and 3.78° respectively (from the SolarSoft STEREO orbit tool). Distance c' was calculated to be 62.26 R$_\odot$ and e (also from the SolarSoft STEREO orbit tool) was 206.83 R$_\odot$.

It is important to note that the cadence of HI-1 is forty minutes, whereas EISCAT observations of IPS sample at 100 Hz and a well-defined spectrum is assembled in about one minute. Small-scale solar wind structures can therefore be observed with IPS. The large-scale view of the HI instruments provides an excellent global context for the IPS results.

By observing the propagation of the SIR, as revealed by the position of the transient once it has become entrained, it can be seen that Venus (also in the HI-2A field-of-view) lies in its path. If the SIR is long lived enough then solar rotation will carry it through the Venusian environment. *In-situ* data from the ASPERA-4 (Barabash *et al.*, 2007) instrument on *Venus Express* (VEX: Svedhem *et al.*, 2007) were also available, as discussed in Section 2.5. We emphasize that no *in-situ* signatures of the transient itself were detected at Venus.

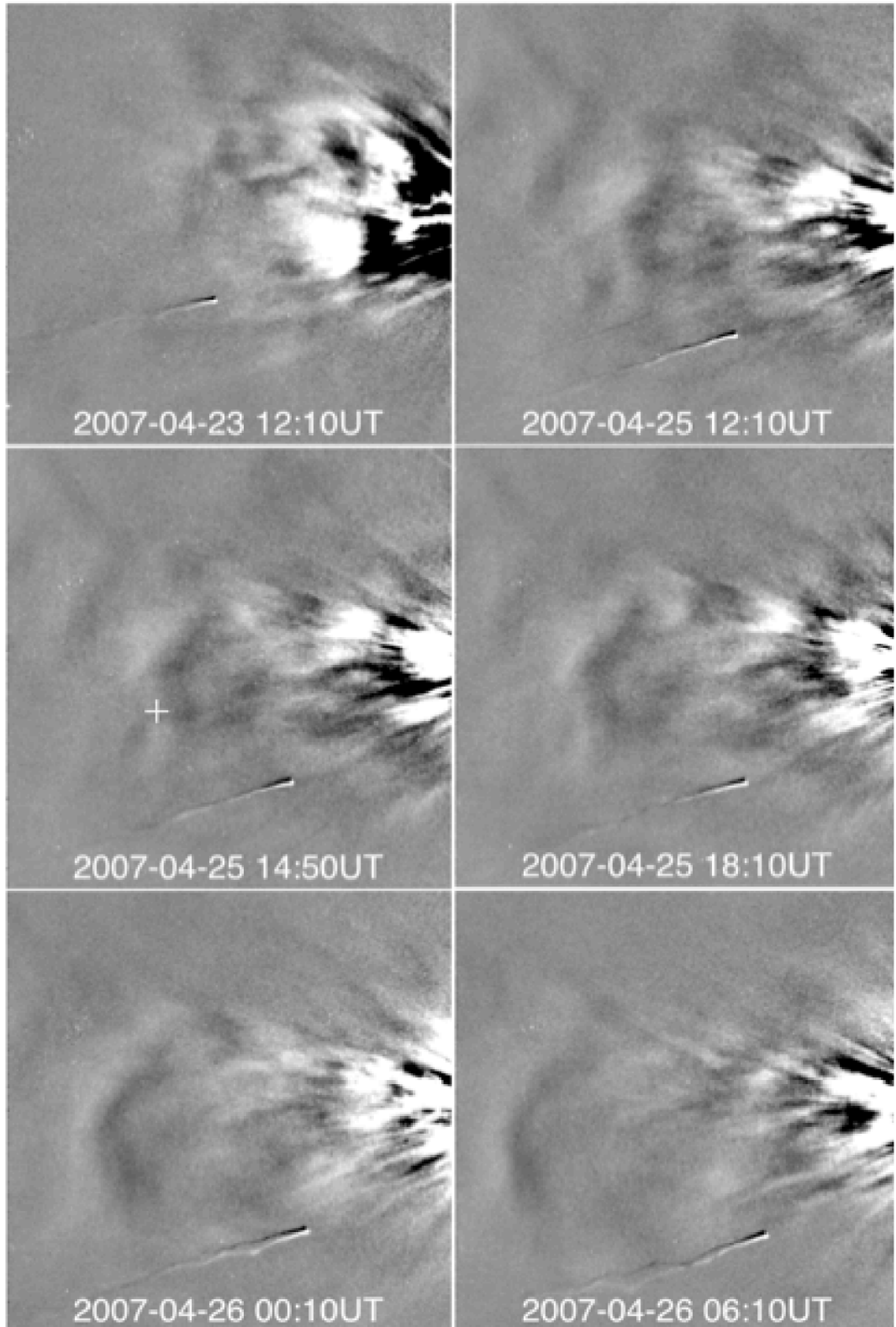

**Figure 10.** A sequence of six running difference images from HI-1A (note the images are not equally spaced in time). The transient appears as a latitudinally extended density enhancement, which crosses P' (white cross), in the third image, at approximately 14:50 UT. This time coincides with the sixty-minute observation of IPS beginning at 14:30 UT. The developing compression region in the last four images increases longitudinally, compressing the transient.

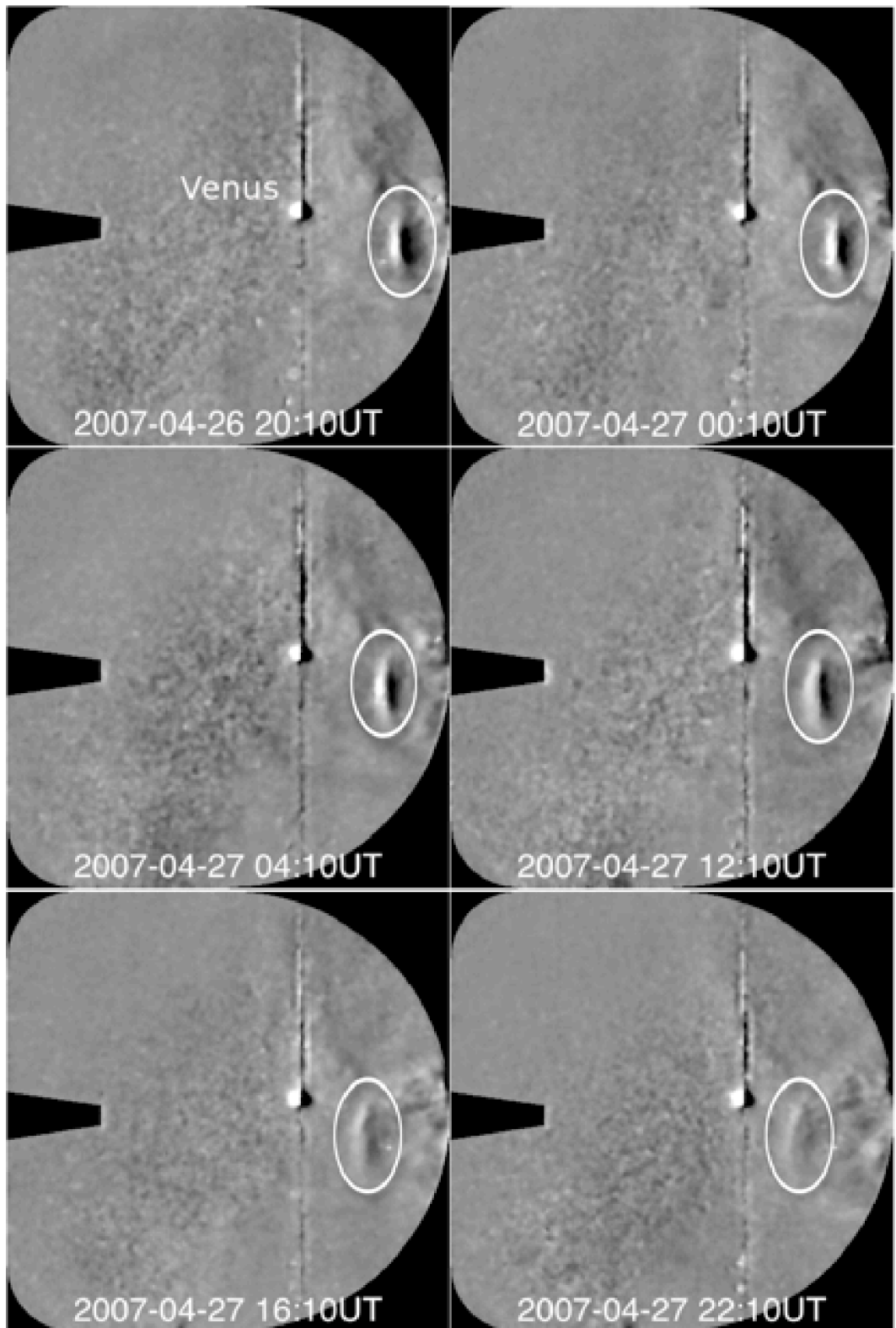

**Figure 11.** A series of six running difference HI-2A images taken over a period of two days immediately following observations of IPS on 25 April. The transient (circled) is now compressed against the leading edge of the SIR. As the material density becomes rarefied with increasing heliocentric distance, and the transient moves away from the HI-2A Thomson sphere, the structure begins to fade from view, as can be seen in the last two images. Venus can be seen to lie in the two-dimensional (2-D) projection path of the SIR. *In-situ* data from *Venus Express* were available from this time and is discussed in Section 2.5.

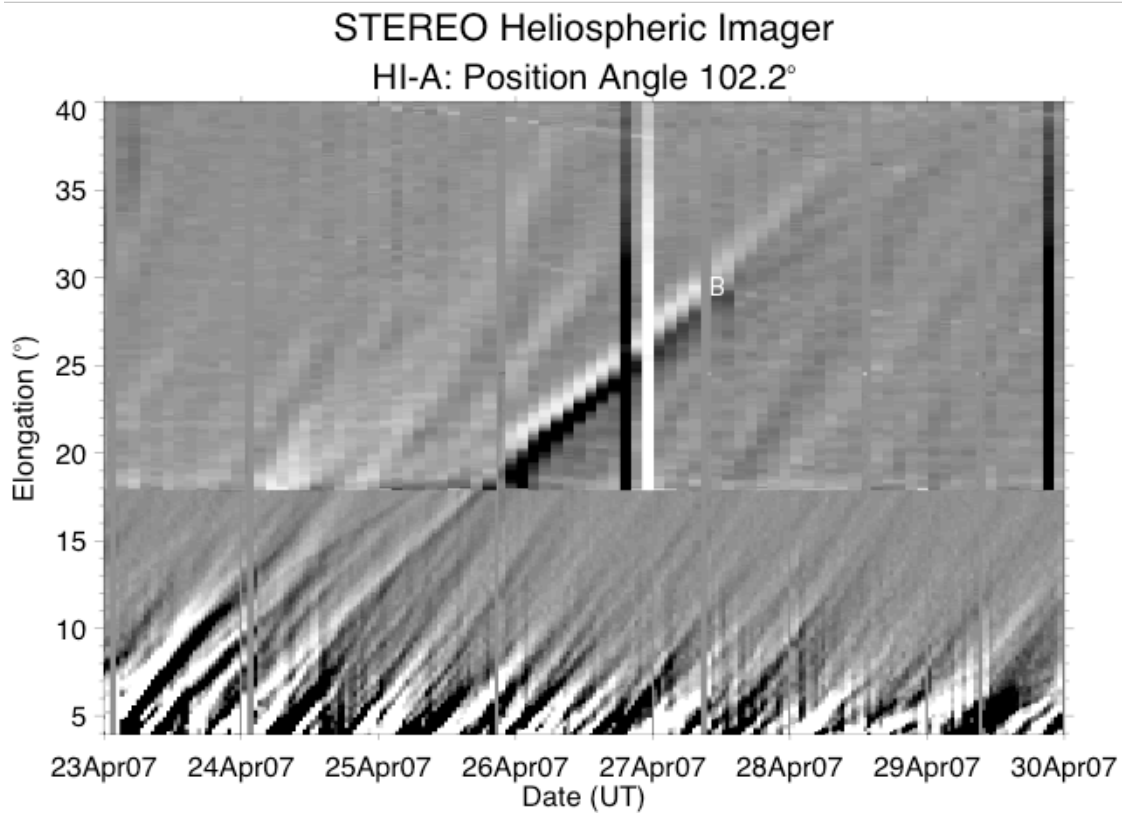

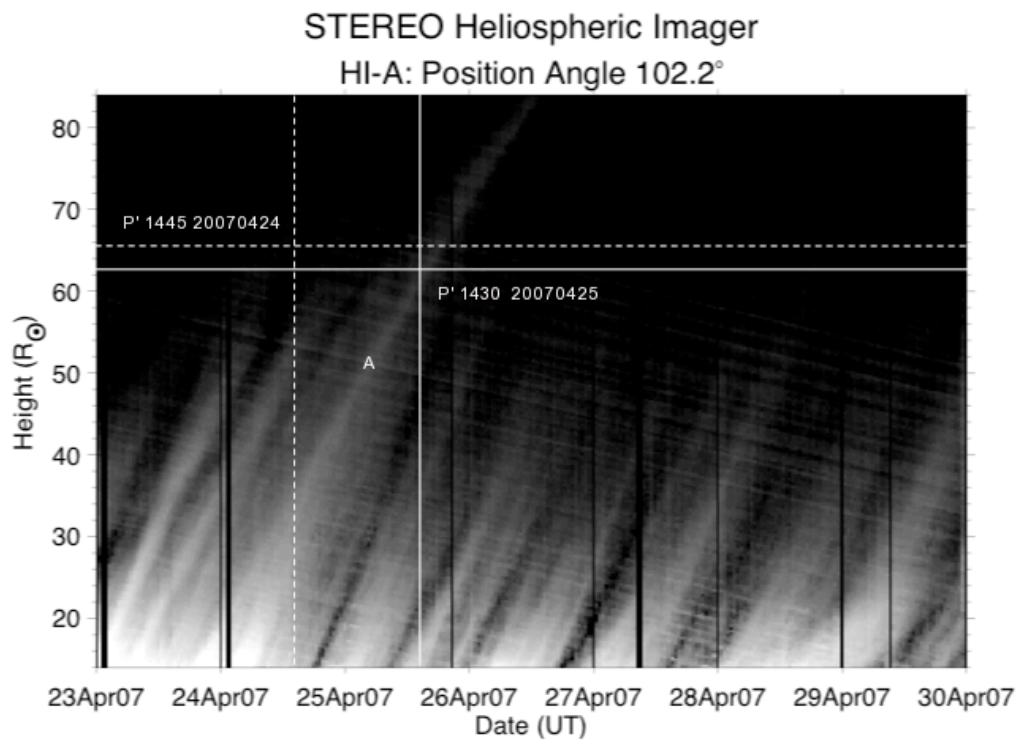

**Figure 12**. The top panel shows a running difference image J-map combining HI-1A and HI-2A data covering the period of interest. The bottom panel shows the same time period but for HI-1A only (and also plotted as a function of POS height not elongation). These plots are produced by combining slices of running difference HI-1A and HI-2A images (top panel) and background subtracted HI-1A images (bottom panel). Feature A is the meso-scale transient with a radial velocity of 273±52.5 km s$^{-1}$, prior to being entrained in the SIR. P' on 24 April (bottom panel) is shown not to overlay any significant white light features. P' on 25 April overlies the transient, giving rise to the stable anti-correlation observed in the cross-correlation functions on this day. Feature B shows the transient propagating through the HI-2A field-of-view.

## 2.4 Modelling the SIR propagation

The propagation of the SIR, though not the entrained transient, was modelled using data from the IPS observations. This we do for two principal reasons. Firstly, we wish to further test the assumption, made in the IPS weak-scattering model, that the identified coronal hole boundary is the likely source of the SIR. Secondly, we wish to predict when the SIR would sweep through the near-Venus environment, and hence be detectable in *Venus Express* ASPERA-4 data.

The Carrington longitude of the boundary region of the coronal hole, as shown in Figure 6, is 290° whilst the Earth lay at a Carrington longitude of 29°. The angle from the Earth-Sunline of the coronal hole boundary is therefore 99°. A SOHO EIT image of the coronal hole is shown in Figure 13.

The model input parameters were the radial velocity of the SIR, as obtained from the IPS weak-scattering model, and the heliographic longitude of the coronal hole boundary, as obtained from the tomographic inversion plot in Figure 6. It is also assumed that the SIR undergoes no significant acceleration after leaving the lower corona, although this is likely to be somewhat inaccurate. The modelled position of the SIR, on 25 April 2007, is presented on a polar plot of the solar system, in Figure 14, showing the positions of the Sun (at the origin), Earth, Venus, and the IPS ray path (black line). The position of Venus is also plotted. For the purposes of the model, it is assumed that Venus, the Earth, STEREO-A, the Sun, and the SIR, lie in the ecliptic plane. If the velocities from the weak-scattering model are accurate, we should find that the arc of the SIR intersects the IPS ray path. The model output plots show the CIR position corresponding to the radial velocity values obtained with the weak-scattering model as $392\pm50$ km s$^{-1}$.

If SIR signatures are detected *in-situ* at Venus then it is important to examine whether they can be connected with the same event being observed by HI and in IPS data. This is done by again modelling the predicted propagation arc of the SIR and its arrival time at Venus, again using velocities from the weak scattering IPS model. SIR signatures in the ASPERA-4 data were first detected on 30 April (see Section 2.5), 5 days after the IPS observations. During this time the fast wind source point had rotated significantly, and so the SIR position was re-modelled as shown in Figure 15 to take account of the rotation. This second output, again a heliocentric polar plot, shows the SIR position on 30 April 2007 at 15:00 UT, by which time the coronal hole boundary had rotated to a heliographic longitude of approximately 30° (Fig. 13).

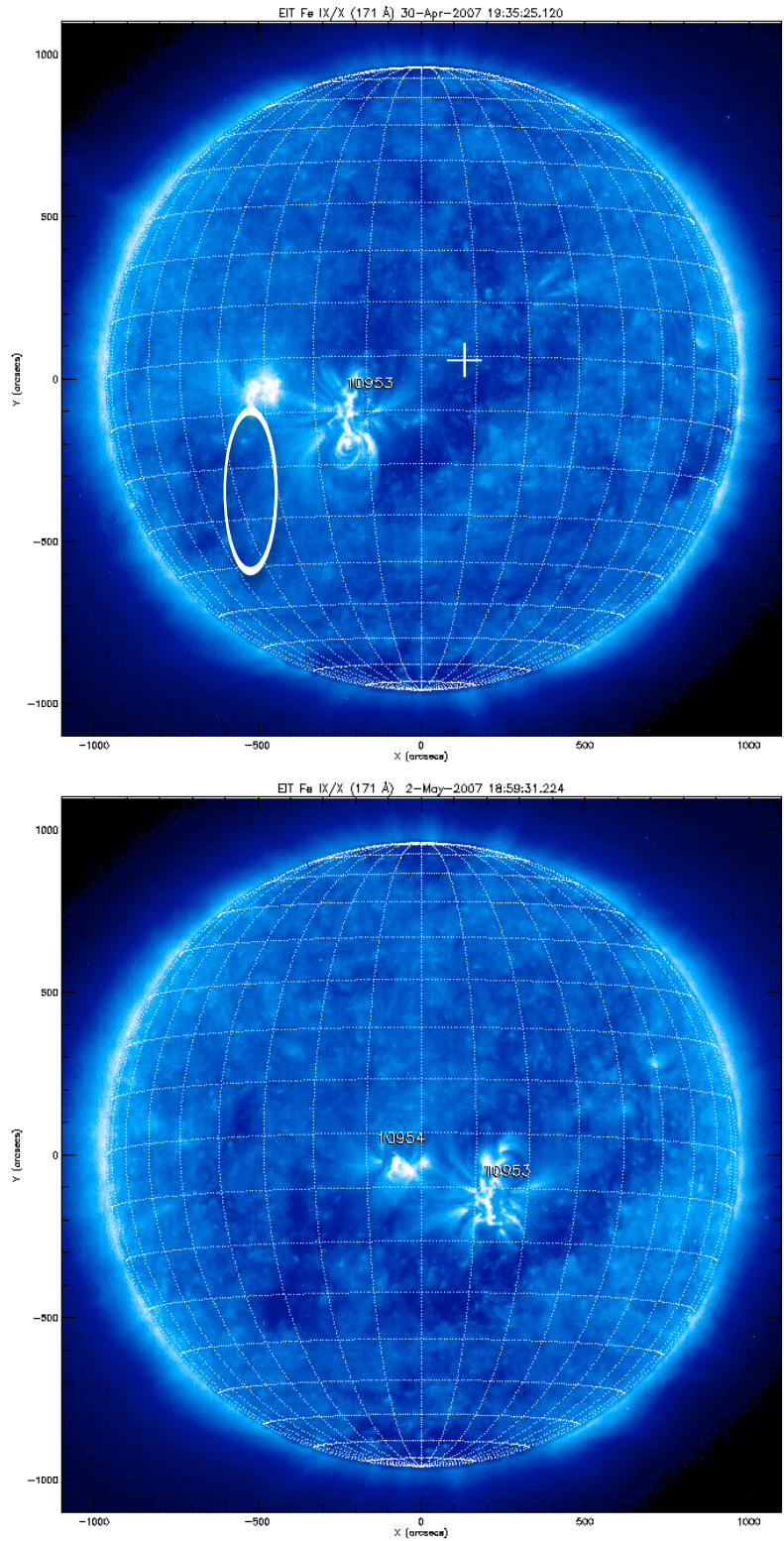

**Figure 13.** The top EIT image is from 30 April 2007. A 10° grid in heliographic latitude and longitude is overlaid with the central meridian at 0°. The coronal hole boundary (circled) is visible just to the left of the substantial active region AR10953, at approximately 30° E. The white cross shows the approximate position of the sub-STEREO-A point. The hole itself is somewhat obscured by the AR10953 so an EIT image from 2 May 2007 is also shown (bottom). In this image the coronal hole is now clearly visible, as it has rotated further into view.

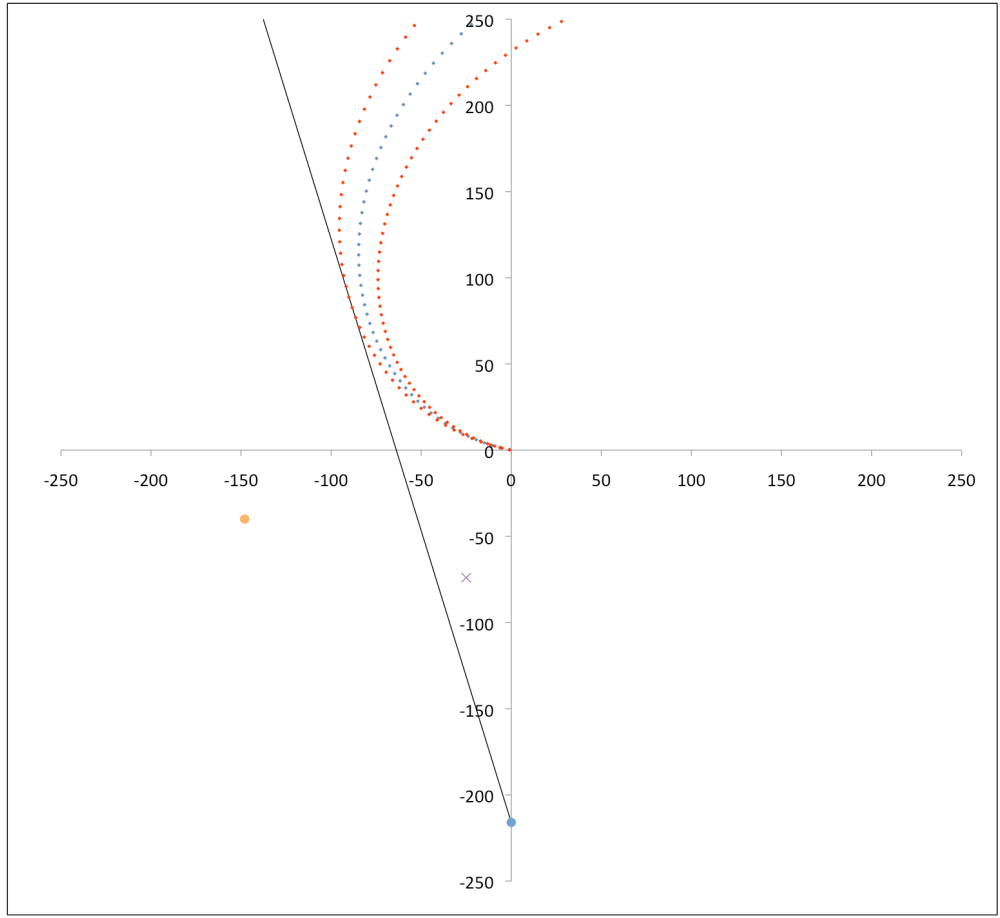

**Figure 14.** This heliocentric polar plot shows the position of the SIR (blue diamond spiral) at 15:00 UT on 25 April 2007, in the ecliptic plane. The propagation velocity used in this model was 392±50 km s$^{-1}$, as obtained from fitting with the IPS weak-scattering model. The inner and outer red diamond spirals correspond to the SIR as modelled for propagation velocities of 342 km s$^{-1}$ and 442 km s$^{-1}$. These correspond to error bounds for the radial velocity as defined by the range of velocities (±50 km s$^{-1}$) across the ray path, used as a parameter in the IPS weak-scattering model. Earth and Venus are the blue and orange spots, respectively, and the axes are in solar radii. The Sun is at the origin and the IPS ray path is shown as the black line. The purple cross shows the position of comet Encke.

There are a number of differences in the model outputs for 15:00 UT on 30 April (Figure 15) as compared with those from 25 April (Figure 14). Firstly, they are given from the viewpoint of STEREO-A, rather than Earth (i.e. STEREO-A is at X=0). The angle of separation between Earth and STEREO-A at this time was 4.19°, so the angle between the sub-STEREO-A point and the coronal hole boundary is ~34°. Secondly, a line-of-sight from STEREO-A is projected from Earth at an elongation of approximately 35°. This corresponds to the elongation at which the SIR-entrained transient directly observed in HI-2A starts to fade from view (Figure 11, bottom right image and the J-map in Figure 12). The third difference is that the Thomson sphere for STEREO-A has also been added. One additional run was also performed for 30 April in which the radial velocity of the SIR is selected so that it passes over Venus, whilst not varying the heliographic longitude of the coronal hole boundary (lower right panel). The velocity required for the modelled SIR to intersect Venus on 30 April was 455 km s$^{-1}$, as compared with the value of 392 km s$^{-1}$ of the SIR obtained from the IPS weak scattering model fitting.

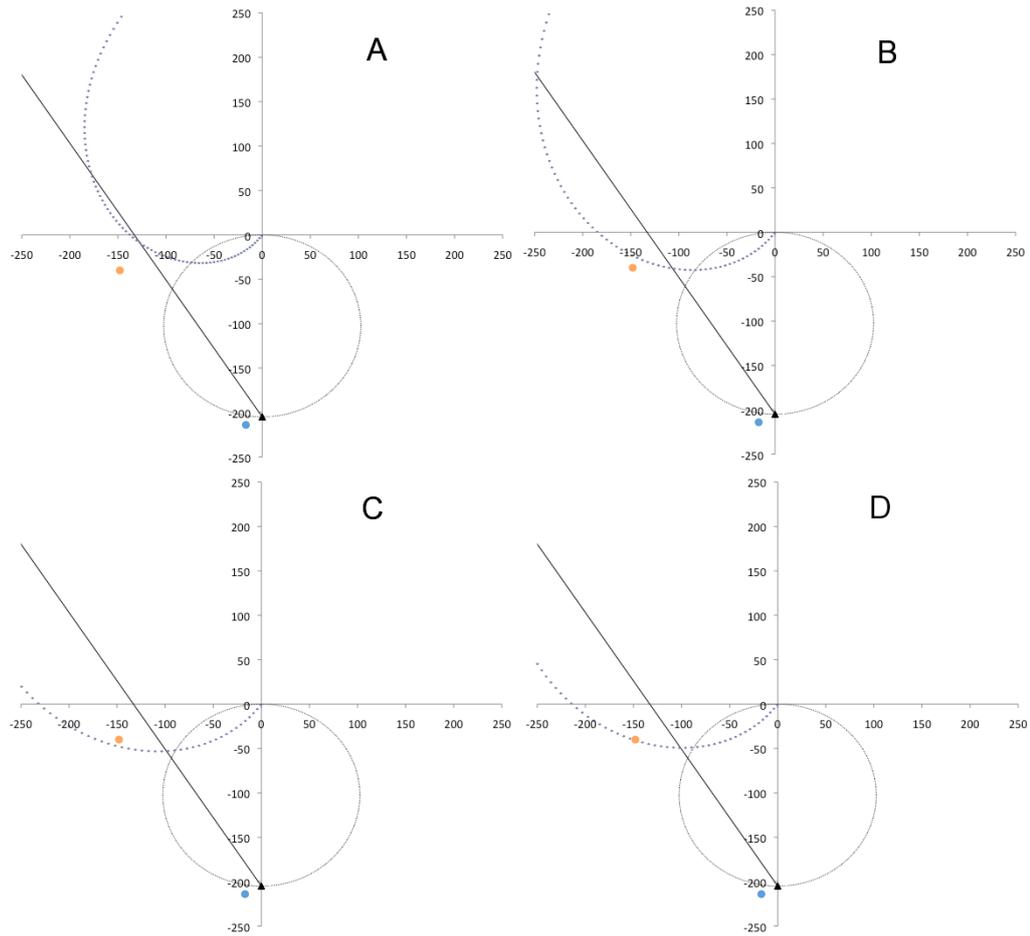

**Figure 15.** A, B, C and D are heliocentric polar plots for the modelled SIR position on 30 April 2007. Earth and Venus are the blue and orange dots respectively and STEREO-A is the black triangle. The SIR is shown as the blue dotted arc. The Thomson sphere for STEREO-A is plotted, as well as a line-of-sight corresponding to 35° elongation, beyond which in all cases it can be seen that viewing geometry for HI becomes increasingly unfavourable. To assess the position of the SIR relative to Venus, a range of propagation velocities were used. In plot A the SIR is modelled with a velocity of 292 km s$^{-1}$, in B at 392 km s$^{-1}$, in C at 492 km s$^{-1}$ and finally in D with a velocity of 455 km s$^{-1}$.

Any increase in Thomson scattered light received by HI-2A due to an increase in the transient density, as it becomes compressed through its interaction with the leading edge of the SIR, is offset by the fact that the SIR-entrained transient is moving away from the Thomson sphere of the Heliospheric Imager. It can be seen in Figure 15 that for elongation angles of greater than 35°, viewing geometry becomes increasingly unfavourable. The plots of the SIR position demonstrate a high probability the SIR will intercept Venus near 30 April 2007. Details of *in-situ* measurements taken at this time by the *Venus Express* spacecraft are discussed in the following section.

## 2.5 Venus Express Data

The *Venus Express* spacecraft is in a highly elliptical orbit about Venus, which takes it through the induced magnetosphere and bow shock and then into the solar wind.

Solar wind ion data from ASPERA-4 from an extended period covering 25 April to 7 May are shown in Figure 16. Only the sixty-minute data slices recorded when VEX is at apoapsis, and hence lying outside of the Venusian magnetosphere, are presented in this panel.

An increase in count rate can be seen on 30 April. In addition, over the next two days a rise in ion energy, which reaches a peak sometime on 3 May, can also be seen. From this point onwards the particle energy then gradually returns to the same level recorded on 29 April. This combination of signatures corresponds to the SIR and trailing fast wind sweeping through the Venusian environment. The VEX ASPERA-4 IMA results from this period are discussed in more detail in an accompanying paper to this topical issue (Whittaker *et al.*, 2010).

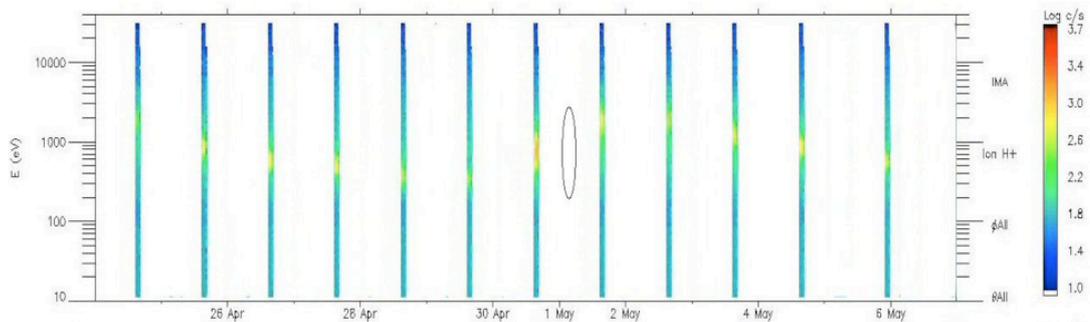

**Figure 16.** The solar wind ion data on 30 April shows a significant enhancement in the count rate and particle energy. Count rate is plotted as a function of time and energy, with colour indicating particle count as per the legend. Every orbit is 24 hours long and each solar wind data slice lasts for 60 minutes, taken when VEX was at apoapsis.

## 3. Discussion

EISCAT observations of IPS using J0318+164 on 21, 22, and 24 April 2007 show significant solar wind transient signatures in the ray path. These are revealed by the presence of strong but short lived (< 30 minutes) anti-correlation in the cross-correlation function, indicative of magnetic field rotations occurring in solar wind density features crossing the ray path.

Rapid changes in $V_{PoS}$ are also recorded over timescales of minutes. For example, on 24 April, the speed increased from 300±5 to 381±11 km s$^{-1}$ over a 20-minute interval beginning at about 15:00 UT (See Figure 4). These changes in speed are attributed to the same transient features that give rise to the field rotations.

The observations of IPS from 25 April are of particular interest as they display significantly enhanced scintillation levels recorded at both the Kiruna and Sodankylä antennas. In observations from the previous days, the scintillation level at Sodankylä did not exceed 0.8 (normalized arbitrary units). However, for the duration of the 25 April observations, the scintillation level did not drop below 1.5. Likewise, the scintillation level at Kiruna on the three observation days prior to 25 April did not exceed 1.5. However, on 25 April it did not drop below 3.5. On both antennas, the scintillation levels were also observed to be much more variable on the 25th. Note that the difference in scintillation levels recorded at each antenna is due to the different signal to noise conditions present at each, however they both experienced a significant increase on the 25th.

The presence of the negative lobes in the IPS cross-correlation functions throughout observations on 25 April indicates the presence of a meso-scale transient feature within the slow solar wind. This structure is clearly seen by HI-1A and HI-2A (Figs. 10, 11 and 12) and bears a strong resemblance to the structure presented by Rouillard *et al.*, (2009). The radial velocity of the transient as quoted on the STEREO HI event list was 273±53 km s$^{-1}$. Further, the propagation direction of the central axis of motion, also obtained from the HI event list, was 89.3±15° with an estimated launch time of 16:28 UT on 23 April (this assumes constant speed back to Sun). It is possible, therefore, that active region AR10953 was the transient source as it lay within 15° of the central axis of motion on 23 April.

Ballistically mapping the ray path for the 25 April observation onto a LASCO tomographic map of the corona (at a height of 2.5 R$_\odot$), showed that a portion of the ray path lying between 35° and 60° sourceward of the P-point was likely to be overlying the boundary between the equatorial coronal hole and the surrounding corona, including active region AR10953. Assuming that the compression region occupies this portion of the ray path, the IPS weak scattering model returned a radial velocity for the SIR compression region of 392±50 km s$^{-1}$.

It is also important here to highlight the fact that IPS sampling is performed on timescales of seconds or less, whereas HI-1 and HI-2 operate with cadences of 40-minutes and 120-minutes, respectively. HI-1 and HI-2 observations therefore provide an excellent global context for the high time resolution IPS observations. IPS remains, thus far, the only remote sensing technique available with the capability to sample solar wind velocity and variability at all heliographic latitudes and at such large heliocentric distances.

Modeling the SIR arrival time at P', in Section 2.4, provides evidence that this velocity value for the SIR is accurate within its quoted errors. It was also accurate enough to predict the arrival time of the SIR at Venus to within 24 hours, albeit with a slight shortfall (Figure 15b). This suggests that the velocity of the SIR was being slightly underestimated from its true value, suggesting that the SIR may have undergone some acceleration after leaving the Sun and/or that the portion of the SIR sampled with IPS observation is not the same as that which is detected *in-situ* at Venus. Nevertheless the discrepancy is small and, as stated, the modeled SIR positions are accurate enough to predict an arrival time at Venus to within 24 hours. This is comparable to the solar wind data measurement time for VEX (60 minutes per 24 hours).

It is clear from Figure 16 that no significant signatures that could be interpreted as being SIR related were observed on 29 April at Venus. However, by 1 May a substantial increase in particle count was detected in the solar wind itself. This was followed over the next several days by a steady rise in particle energies as the rarefaction region of the SIR (containing fast solar wind) swept through the Venusian neighbourhood. There is no obvious indication of the transient in this data, which may be consistent either with the transient missing Venus altogether, or, less probably, that the transient material has become indistinguishable from SIR material.

## 4. Conclusions

The results presented in this paper provide evidence for the presence of small-scale transients in the slow solar wind on 21, 22 and 24 April 2009, associated with regions of significant rotation in the interplanetary magnetic field. These transients have spatial scales of < 400000 km. The presence of rapid variations in magnetic field orientation associated with these features means that they may affect the plasma environment of solar system objects on short timescales. We suggest these features may be the interplanetary counterparts of pixel brightenings reported in coronagraph data by Tappin, Simnett, and Lyons (1999).

The rapid time variation in solar wind $V_{PoS}$ and interplanetary magnetic field orientation caused by these features (certainly on timescales of ~10 minutes) could not have been resolved at the heliocentric distances covered without the rapid sampling available from EISCAT observations of IPS, while STEREO HI observations made it possible to associate these small-scale structures with larger slow wind features. Without the combination of results from the two techniques, it would have been very difficult to determine the true nature of these rapid variations.

A meso-scale transient, with a minimum scale size of 1180000 km, was detected in observations of IPS on 25 April 2009. The IPS observations suggested that it lay within the slow solar wind close up-rotation from the compression region of the SIR, and had a possible origin in active region AR10953. This larger meso-scale transient discussed here may represent an example of the interplanetary counterpart to the "Sheeley blobs," seen closer to the Sun in LASCO observations (Sheeley *et al*., 1997).

We suggest that this same transient was imaged by HI on STEREO-A. The velocity of the transient, as obtained from HI data, was 273±53 km s$^{-1}$, and the modeled velocity of the SIR compression region was 392±50 km s$^{-1}$ (from fitting IPS data to the weak-scattering model). As in the cases discussed by Russell *et al*. (2009), no evidence for a shock front was found, and only a velocity gradient was present across the compression region of the SIR. The velocity quoted here will thus have contributions from across this gradient and represents an average radial velocity for the SIR.

The slow wind-like velocity of the transient as observed by HI, and our unsuccessful attempts to model the transient separately from the slow wind, make it very likely that the transient, initially travelling with the slow solar wind, was subsequently entrained by the compression region of the SIR. This is consistent with the longitudinal compression of the structure in STEREO HI-2A images at larger distances from the Sun. We therefore propose that the meso-scale structure detected in IPS data on 25 April 2007 originated as a slow wind feature (possibly from AR10953), was still within the slow wind at the time it crossed the IPS ray path and was subsequently swept up by the SIR.

These observations support the interpretation of a similar structure discussed by Rouillard *et al*. (2009). The difference between the two studies is that here we sample the transient with IPS at an earlier stage in its evolution, prior to it becoming entrained. The transient and the SIR were detected as separate structures at the time of the observations of IPS. Using the position of the ray path as a constraint, we can therefore state that transient entrainment occurred no closer to the Sun than 62.4 $R_\odot$ and not before 14:30 UT on 25 April.

The dynamics of the tail of comet 2P / Encke in the days following the CME-plasma tail disconnection event (Vourlidas *et al*., 2007) were used by Clover *et al*. (2010) to probe the velocity of solar wind tranisents over the period of interest here. The authors reported highly variable solar wind velocities in the locality of the plasma tail, ranging from 200 km s$^{-1}$ to 580 km s$^{-1}$. Comet 2P / Encke lay in close proximity to the IPS ray path throughout observations on 21 - 24 April. Based on the comparable velocities and close proximity, we suggest that similar transient structures were giving rise both to the anti-correlation signatures observed in IPS data and the motions of the plasma tail.

2P / Encke was near perihelion (0.34 AU) at the time of this study and hence at a comparable heliocentric distance with the IPS P-point. The IPS results reported here support the conclusions of Clover *et al*. (2010) that solar wind velocities at near-Sun distances are highly variable.

No obvious signatures of the meso-scale transient were measured *in-situ* at Venus. However the SIR, that it subsequently became entrained by, did arrive at Venus on 30 April 2007. The recorded arrival time of the SIR in ASPERA-4 data agreed with the modeled propagation velocity to within 24 hours. Results discussed here, and by Rouillard *et al*. (2009), represent a significant development in our ability to resolve and interpret complex solar wind features.

## Acknowledgements

We would like to thank the director and staff of EISCAT, the STEREO SECCHI consortium and the *Venus Express* ASPERA-4 team for the data used in this paper. The STEREO|SECCHI data used here are produced by an international consortium of the Naval Research Laboratory (USA), Lockheed Martin Solar and Astrophysics Lab (USA), NASA Goddard Space Flight Center (USA), Rutherford Appleton Laboratory (UK), University of Birmingham (UK), Max-Planck- Institut für Sonnensystemforschung (Germany), Centre Spatiale de Liège (Belgium), Institut d'Optique Théorique et Appliquée (France), Institut d'Astrophysique Spatiale (France).

EISCAT is supported by the scientific research councils of China, Finland, France, Germany, Japan, Norway, Sweden and the UK. The STEREO spacecraft is a NASA mission and the Heliospheric Imagers were developed by a consortium including Birmingham University and Rutherford-Appleton Laboratory (UK), the Naval Research Laboratory (US) and Centre Spatial de Liege (Belgium). *Venus Express* is managed by the European Space Agency.

We would also like to thank W.A. Coles and B.J. Rickett of University of California, San Diego for valuable advice and discussions during the development of our current IPS analysis programs.

Four of us (GDD, APR, ICW, RAF) were supported by the Science and Technology Facilities Council of the UK during the period when this work was carried out.

# References

Altschuler, M. D., Newkirk, Jr. G.: 1969, Magnetic fields and the structure of the solar corona, *Solar Phys.*, **9(1)**, 131-149.

Armstrong, J.W., Coles, W.A.: 1972, Analysis of 3-station interplanetary scintillation data, *J. Geophys. Res.*, **77(25)**, 4602-4610.

Asai, K., Kojima, M., Tokumaru, M., Yokobe, A., Jackson, B. V., Hick, P. L., Manoharan, P. K.: 1998, Heliospheric tomography using interplanetary scintillation observations. III - Correlation between speed and electron density fluctuations in the solar wind, *J. Geophys. Res.*, **103**, 1991-.

Barabash, S., Sauvaud, J.J., Gunell, H., Anderson, H., Grigoriev, A., Brinkfeldt, K., Holmström,M., Lundin, R., Yamauchi, M., Asamura, K., *et al*.: 2007, The analyser of space plasmas and energetic atoms (ASPERA-4) for the *Venus Express* mission, *Plan. Space Sci.*, **55**, 1772-1792.

Bisi, M. M., Fallows, R. A., Breen, A. R., Habbal, S. R., Jones, R. A.: 2007, Large-scale structure of the fast solar wind, *J. Geophys. Res.*, **112**(A6), A06101.

Bisi, M.M., Fallows, R.A., Breen, A.R., O'Neill, I. J.: 2010, Interplanetary Scintillation Observations of Stream Interaction Regions in the Solar Wind, *Solar Phys.*, **261**, 149-172.

Borovsky, J. E.: 2006, Eddy viscosity and flow properties of the solar wind: Co-rotating interaction regions, coronal-mass-ejection sheaths, and solar-wind/magnetosphere coupling, *Physics of Plasmas,* **13***,* 056505.

Borovsky, J. E.: 2008, Flux tube texture of the solar wind: Strands of the magnetic carpet at 1 AU?, *J. Geophys. Res.*, **113**(A8), A08110

Bourgois, G.: 1972, Study of the solar wind using the power spectrum of interplanetary scintillation of radio sources, *Astron. Astrophys.*, **19**, 200-206.

Bourgois, G., Daign, G., Coles, W.A., Silen, J., Turenen, T., Williams, P.J.S.: 1985, Measurements of the solar wind with EISCAT, *Astron. Astrophys.*, **144**, 452-462.

Breen, A. R., Coles, W. A., Grall, R.R., Klinglesmith, M.T., Løvhaug, U.-P., Markkanen, J. *et al*.: 1996, EISCAT measurements of interplanetary scintillation, *J. Atmos. Terr. Phys.,* **58**, 507-519.

Breen, A. R., Coles, W. A., Grall, R.R., Klinglesmith, M.T., Markkanen, J., Moran, P.J., Tegid, B., Williams, P.J.S.: 1996, EISCAT measurements of the solar wind, *Ann. Geophys.*,**14**, 1235-1245.

Breen, A.R., Moran, P.J., Varley, C.A., Wilkinson, W.P., Williams, P.J.S., Coles, W.A., Lecinski A., Markkanen, J.: 1998, Interplanetary scintillation observations of interaction regions in the solar wind, *Ann. Geophys.*, **16**, 1265-1282

Breen, A.R., Mikic, Z., Linker, J.A., Lazarus, A.J., Thompson, B.J., Moran, P.J., Varley, C.A., Williams, P.J.S., Biesecker, D.A, Lecinski, A.: 1999, Interplanetary scintillation measurements of the solar wind during Whole Sun Month: linking coronal and *in-situ* observations, *J. Geophys. Res.*, **104**, 9847-9870.

Breen, A.R., Fallows, R.A., Bisi, M.M., Thomasson, P., Jordan, C.A., Wannberg, G., Jones, R.A.: 2006, Extremely long baseline interplanetary scintillation measurements of solar wind velocity, *J. Geophys. Res.*, **111**, A08104.

Bourgois, G., Daigne, G., Coles, W. A., Silen, J., Turunen, T., Williams, P. J. S.: 1985, Measurements of the solar wind velocity with EISCAT, *Astron. Astrophys.*, **144**(2), 452-462.

Canals, A.: 2002, Interplanetary scintillation studies of the solar wind during the rising phase of the solar cycle, Ph.D. thesis, University of Wales, Aberystwyth.

Canals, A., Breen, A.R., Moran, P.J., Ofman, L.: 2002, Estimating random transverse velocities in the fast solar wind from EISCAT Interplanetary scintillation measurements, *Ann. Geophys.*, **20**, 1265-1277.

Clover J.M., Jackson, B.V., Bufington, A., Hick, P.P., Bisi, M. M.:2010, Solar wind speed inferred from cometary plasma tails using observations from STEREO HI-1, *Ap. J.*, **713**, 394-397.

Coles, W.A.: 1995, Interplanetary scintillation observations of the high latitude solar wind, *Space Sci. Rev.*, **72**, 211-222.

Coles, W.A.: 1996, A bimodal model of the solar wind, *Astrophys. Space. Sci.*, **243(1)**, 87-96.

Davies, J.A., Harrison, R.A., Rouillard, A.P., Sheeley, N.R., Perry, C.H., Bewsher, D., Davis, C. J., Eyles, C. J., Crothers, S. R., Brown, D. S.: 2009, A synoptic view of solar transient evolution in the inner heliosphere using the Heliospheric Imagers on STEREO, *Geophys. Res. Letts.*, **36**, L02102.

Dorrian, G.D., Breen, A.R., Brown, D.S., Davies, J. A., Fallows, R. A., Rouillard, A. P.: 2008, Simultaneous Interplanetary Scintillation and Heliospheric Imager observations of a coronal mass ejection, *Geophys. Res. Letts.*, **35**, L24104.

Eyles, C.J., Simnett, G.M., Cooke, M.P., Jackson, B.V., Buffington, A., Hick, P.P., Waltham, N.R, King, J. M., Anderson, P. A., Holladay, P. E.: 2003, The solar mass ejection imager (SMEI), *Solar Phys.*, **217**, 319-347.

Eyles C.J., Harrison, R.A., Davis, C.J., Waltham, N.R., Shaughnessy, B.M., Mapson-Menard, H.C.A., Bewsher, D., Crothers, S. R., Davies, J. A., Simnett, G. M., *et al*: 2009, The Heliospheric Imagers Onboard the STEREO Mission, *Solar Phys.*, **254**, 387-445.

Fallows, R. A., Breen, A. R., Dorrian, G.D.: 2008, Developments in the use of EISCAT for interplanetary scintillation, *Ann. Geophys.*, **26**, 2229-2236.

Fallows, R. A., Breen, A. R., Bisi, M.M., Jones, R.A., Wannberg, G.: 2006, Dual-frequency interplanetary scintillation observations of the solar wind, *Geophys, Res. Letts*, **33**, L11106.

Gosling, J. T., Asbridge J. R., Bame, S. J., Feldman, W. C.: 1978, Solar Wind Stream Interfaces, *J. Geophys. Res.*, **83** (A4), 1401-1412

Grall, R.R., Coles, W.A., Klinglesmith, M.T., Breen, A.R., Williams, P.J.S., Markkanen, J., Esser, R.: 1996, Rapid Acceleration of the polar solar wind, *Nature*, **379**, 429-432.

Grall, R. R., Coles, W. A., Spangler, S. R., Sakuri, T., Harmon, J. K.: 1997, Observations of field-aligned density microstructure near the Sun, *J. Geophys. Res.*, **102**(A1), 263-273.

Harrison R.A., Davies, J.A., Rouillard, A.P., Davis, C.J., Eyles, C.J., Bewsher, D., Crothers, S.R. Russell, R. A., Sheeley, N. R., Vourlidas, A., *et al*.: 2009, Two years of the STEREO Heliospheric Imagers : Invited review, *Solar Phys.*, **256**, 219-237.

Hewish, A.P., Scott, P., Wills, D.: 1964, Interplanetary scintillation of small diameter radio sources, *Nature*, **203**, 1214-1217

Howard R. A., Moses, J.D., Vourlidas, A., Newmark, J.S., Socker, D.G., Plunkett, S.P., Korendyke, C. M., Cook, J. W., Hurley, A., Davila, J. M., *et al*.: 2008, Sun Earth Connection Coronal and Heliospheric Investigation (SECCHI), *Space. Sci. Rev.*, **136**, 67-115.

Jackson B. V.: 1985, Imaging of Coronal Mass Ejections by the Helios Spacecraft, *Solar Phys.*, **100**, 563-574.

Jackson, B.V., Buffington, A., Hick, P.P., Altrock, R.C., Figueroa, S., Holladay, P.E., Johnston, J. C., Kahler, S. W., Mozer, J. B., Price, S., *et al*.: 2004, The solar mass-ejection imager (SMEI) mission, *Solar Phys.*, **225**, 177-207.

Jones, R.A., Breen, A.R., Fallows, R.A., Canals, A., Bisi, M.M.: 2007, Interaction between coronal mass ejections and the solar wind, *J. Geophys. Res.*, **112**, A08107.

Kaiser M. L., 2005, The STEREO mission: an overview, *Adv. in Space Res.*, **36**(8), 1483-1488.
Kaiser M. L., Kucera, T. A., Davila, J. M., St Cyr, O. C., Guhathakurta, M., Christian, E.: 2008, The STEREO Mission: An Introduction, *Space Sci. Rev.*, **136**(1-4), 5-16.

Klinglesmith, M.T.: 1997, The polar solar wind from 2.5 to 40 solar radii; Results of intensity scintillation measurements, Ph.D. thesis, University of California, San Diego.

Klinglesmith, M.T., Grall, R. R., Coles, W. A.: 1996, 933 MHz IPS Velocity Measurements at EISCAT, AIP Solar Wind 8 Conference Proceedings, **382**, 180-183.

Marsch, E., Muhlhauser, K.H., Schwenn, R., Rosenbauer, H., Pilipp, W., Neubauer, F.M.: 1982, Solar wind protons: 3 dimensional velocity distributions and derived plasma parameters between 0.3 AU and 1 AU, *J. Geophys. Res.*, **87**, 52-72.

Morgan, H., Habbal, S.R., Lugaz, N.: 2009, Mapping the structure of the corona using Fourier back projection tomography, *Ap. J.*, **690**, 1119-1129.

Neugebauer, M., Snyder, C.W.: 1966, Mariner 2 observations of the solar wind, *J. geophys. Res.*, **71**, 4469-4484.

Rishbeth, H., Williams, P.J.S.: 1985, The EISCAT ionospheric radar: The system and its early results, *Mon. not. R. Astr. Soc.*, **26**, 478-512.

Rouillard A.P., Davies, J.A., Forsyth, R.J., Rees, A., Davis, C.J., Harrison, R.A., Lockwood, M., Bewsher, D., Crothers, S. R., Eyles, C. J., *et al*.: 2008, First imaging of corotating interaction regions using the STEREO spacecraft, *J. Geophys. Res.*, **35**, L10110.

Rouillard, A.P., Savani, N.P., Davies, J.A., Lavraud, B., Forsyth, R.J., Morley, S.K., Opitz, A., Sheeley, N. R., Burlaga, L. F., Sauvaud, J. -A., *et al*.: 2009, A multispacecraft analysis of a small-scale transient entrained by solar wind streams, *Solar Phys.*, **256**, 307-326.

Russell, C. T., Jian, L. K., Blanco Cano, X., Luhmann, J. G., Zhang, T. L.: 2009, STEREO observations of shock formation in the solar wind, *Geophys. Res. Letts.* **36**, L02103

Schatten, K. H., Wilcox, J. M., Ness, N. F.: 1969, A model of interplanetary and coronal magnetic fields, *Solar Phys.*, **6(3)**, 442-455

Sheeley, N.R., Wang, Y.-M., Hawley, S.H., Brueckner, G.E., Dere, K.P., Howard, R.A., Koomen, M. J., Korendyke, C. M., Michels, D. J., Paswaters, S. E., *et al*.: 1997, Measurements of flow velocities in the corona between 2 and 30 R-circle dot, *Ap. J.*, **484**. 472-.

Slavin, J.A., Acuna, M.H., Anderson, B.J, Barabash, S., Benna, M., Boardsen, S.A., Gloeckler, G., Gold, R. E., Ho, G. C., Korth, H., *et al*.: 2009, MESSENGER and *Venus Express* observations of the solar wind interaction with Venus, *Geophys. Res. Letts.*, **36**, L09106.

Svedhem, H., Titov, D. V., McCoy, D., Lebreton, J.-P., Barabash, S., Bertaux, J.-L., Drossart, P., Formisano, V., Häusler, B., Korablev, O., *et al*.: 2007, *Venus Express*—The first European mission to Venus, *Planetary and Space Science*, **55**(12), 1636-1652.

Tappin, S. J., Simnett, G. M., Lyons, M. A.: 1999, A determination of the outflow speeds in the low solar wind, *Astronomy and Astrophysics*, **350**, 302-309.

Vourlidas, A., Howard, R.A.: 2006, The proper treatment of coronal mass ejection brightness: A new methodology and implications for observations, *Ap. J.*, **642**, 1216-1221.

Vourlidas, A., Davis, C.J., Eyles, C.J., Crothers, S.R., Harrison, R.A., Howard, R.A., Moses, D. J., Socker, D. G.: 2007, First direct observation of the interaction between a comet and a coronal mass ejection leading to a complete plasma tail disconnection, *Ap. J. Letts.*, **668**, L79-L82.

Wang, Y.-M, Sheeley, N. R.: 1992, On potential field models of the solar corona, *Ap. J*, **392(1)**, 310-319.

Wannberg, G., Vanhein, L. -G., Westman, A., Breen, A. Williams, P. J. S.: 2002, The new 1420 MHz dual-polarisation interplanetary scintillation (IPS) facility at EISCAT, *Proceedings of the Union of Radio Scientists* (URSI) 2002

Watanabe, T., Kakinuma, T., Kojima, M., Shibasak, K.: 1973, Solar wind disturbances detected by scintillation of radio sources in early August 1972, *J. Geophys. Res.,* **78( 34)** ,  8364-8366.

Whittaker, I. C., Dorrian, G. D., Breen, A.R., Grande, M.: 2009, *In-situ* observations of a co-rotating interaction region at Venus identified by IPS and STEREO, *Solar Phys.*, doi: 10.1007/s11207-010-9608-2